\documentclass[aps,prb,showpacs,twocolumn,superscriptaddress]{revtex4-1}           

\usepackage{times,xspace}
\usepackage{amsbsy,amssymb,amsmath,bm}
\usepackage{graphicx,color,epsfig,rotate}

\begin{document}

\title{Momentum dependence in K-edge resonant inelastic x-ray scattering and \\*
       its application to screening dynamics in CE-phase La$_{0.5}$Sr$_{1.5}$MnO$_4$}

\author{T. F. Seman}
    \email{Current address: Department of Physics, Northern Illinois University, De Kalb, Illinois 60115, USA
   and Advanced Photon Source, Argonne National Laboratory, Argonne, Illinois 60439, USA}
    \affiliation{Department of Physics, New Jersey Institute of Technology, Newark, New Jersey 07102, USA\\}
\author{X. Liu}
    \affiliation{Condensed Matter Physics and Materials Science Department, Brookhaven National Laboratory, Upton, New York 11973, USA\\}
    \affiliation{Beijing National Laboratory for Condensed Matter Physics, and Institute of Physics, Chinese Academy of Sciences, Beijing 100190, China\\}
\author{J. P. Hill}
    \affiliation{Condensed Matter Physics and Materials Science Department, Brookhaven National Laboratory, Upton, New York 11973, USA\\}
\author{M. van Veenendaal}
    \email{veenendaal@niu.edu}
     \affiliation{Department of Physics, Northern Illinois University, DeKalb, Illinois 60115, USA\\}
     \affiliation{Advanced Photon Source, Argonne National Laboratory, Argonne, Illinois 60439, USA\\}
\author{K. H. Ahn}
    \email{kenahn@njit.edu}
    \affiliation{Department of Physics, New Jersey Institute of Technology, Newark, New Jersey 07102, USA\\}


\begin{abstract}
    We present a formula
    for the calculation of K-edge resonant inelastic x-ray scattering
    on transition metal compounds, 
    based on a local interaction between the valence shell electrons and the $1s$ core hole.
    Extending a previous result, we include explicit momentum dependence and
    a basis with multiple core-hole sites. We apply this formula to a single-layered charge, orbital and
    spin ordered manganite, La$_{0.5}$Sr$_{1.5}$MnO$_4$, and obtain good agreement with experimental data,
    in particular with regards to the large variation of the intensity with momentum. We find that
    the screening in La$_{0.5}$Sr$_{1.5}$MnO$_4$ is highly localized around the core-hole site and
    demonstrate the potential of K-edge resonant inelastic x-ray scattering as a probe of screening
    dynamics in materials.
\end{abstract}


\maketitle

\section{Introduction}
There has been a great interest recently in K-edge resonant inelastic x-ray scattering
(RIXS),~\cite{Hasan00,Devereaux03,Ishii04,Grenier05,Markiewicz06,Takahashi07,Tyson07,Ament11}
particularly in transition metal oxides, because of its unique advantages over other probes. In this
spectroscopy, hard x-rays with energies of the order of 10~keV excite transition-metal $1s$ electrons
into empty $4p$ levels, which decay back to the $1s$ levels. In addition to the elastic process,
inelastic processes occur that result in low energy excitations of the order of 1~eV near the Fermi
energy, the cross section of which is enhanced by the resonant condition. The K-edge RIXS spectrum
provides information on the momentum dependence of the excitations, is sensitive to the bulk properties
because of the high energy of hard x-rays, and directly probes valance-shell excitations because
there is no core hole in the final states. Since early studies on nickel-based
compounds,~\cite{Kao96,Platzman98} K-edge RIXS has been a useful probe for novel excitations in
transition metal oxides, in particular, high-T$_c$ cuprates.~\cite{Hill98,Abbamonte99}

Theoretically, it has been proposed that the K-edge RIXS spectrum reflects different aspects of
the electronic structure depending on the size of the core-hole potential, $U_{\rm core}$, between the $1s$
core hole and the $3d$ electrons, relative to the $3d$ band width.~\cite{Ament11} In the weak or strong limit
of $U_{\rm core}$, the ultra-short core-hole lifetime expansion is applicable~\cite{vandenBrink06} and
it has been shown that the K-edge RIXS spectrum corresponds to the dynamic structure factor,
$S({\bf q},\omega)$. Some experimental results indeed show K-edge RIXS spectra similar to $S({\bf q},\omega)$,
but others show deviations.~\cite{Kim09} In the intermediate case of $U_{\rm core}$, numerical calculations
show asymmetric electron-hole excitations and that the RIXS spectrum is substantially modified from
$S({\bf q},\omega)$.~\cite{Ahn09}

One of the main conclusions of Ref.~\onlinecite{Ahn09} is that the K-edge RIXS intensity for transition metal
oxides essentially represents the dynamics of electrons near the Fermi energy, which screen the $1s$ core
hole created by the x-ray.~\cite{Ahn09,Semba08} Tuning the incoming x-ray energy to the absorption edge allows
an approximation in which the sum over the intermediate states is replaced with the single lowest-energy
intermediate state. The study further showed that expanding K-edge RIXS intensity according to the number
of final-state electron-hole pairs is a fast-converging expansion where the one-electron-hole-pair states
dominate, particularly for insulators. The calculation further shows that the electron excitation is from
the unoccupied band throughout entire first Brillouin zone, reflecting the localized nature of the core-hole
screening by electrons in real space. In contrast, the hole excitations are mostly from occupied states close
to the gap to minimize the kinetic energy, particularly when the gap energy is smaller than the band width.

In Ref.~\onlinecite{Ahn09}, the focus was on the energy dependence of electron-hole excitations and the case
of one core-hole site per unit cell. The momentum dependence of the RIXS spectrum and the possibility of
multiple core-hole sites within a unit cell were not considered explicitly. In the current paper, we derive
a formula that includes the full momentum dependence as well as multiple core-hole sites within a unit cell
in the tight-binding approach. The formula is expressed in terms of the intermediate states with a completely
localized $1s$ core hole, and we show that the RIXS spectrum in reciprocal space can be readily compared with
the screening cloud in real space.

As a specific example, we calculate the K-edge RIXS spectrum for La$_{0.5}$Sr$_{1.5}$MnO$_4$ 
and make a comparison with experimental results. 
This material has a layered perovskite structure, which includes two-dimensional MnO$_2$
planes with eight Mn sites per unit cell in the low-temperature spin, orbital, charge, and structure ordered
state.~\cite{Sternlieb96,Moritomo95,Bao96} Experimental results show a dramatic variation of the RIXS intensity
in reciprocal space in spite of the fact that there is almost no change in the peak energy of the energy-loss
feature.~\cite{Liu13} We find good agreement between theory and experiment. By varying the parameter values,
we find a correlation between the variation of the K-edge RIXS spectrum in reciprocal space and the size and
shape of the screening cloud in real space. We further examine the periodicity of the K-edge RIXS
spectrum.~\cite{Kim07}

The paper is organized as follows. Section~\ref{sec:rixs-formula} presents the derivation of the K-edge RIXS
formula in the limit of a completely localized $1s$ core hole. We present the experimental results and
the theoretical model for La$_{0.5}$Sr$_{1.5}$MnO$_4$ in Sec.~\ref{sec:exper} and Sec.~\ref{sec:tight-bind},
respectively. Section~\ref{sec:results} presents the results of our calculations and comparison with
experimental results. Section~\ref{sec:discussion} includes further discussion on our results and
Sec.~\ref{sec:summary} summarizes. Appendix~\ref{sec:app-rixs} shows details of the K-edge RIXS formula
derivation. Appendix~\ref{sec:app-H-itok} shows the RIXS formula in terms of eigenstates with and without
the core hole. Appendix~\ref{sec:app-hamilt} includes the expression of the tight-binding Hamiltonian for
La$_{0.5}$Sr$_{1.5}$MnO$_4$ in reciprocal space. We discuss the actual electron numbers at nominal
``Mn$^{3+}$" and ``Mn$^{4+}$" sites in La$_{0.5}$Sr$_{1.5}$MnO$_4$ in Appendix~\ref{sec:app-eh-number}.
The programs used for our calculations are available online.~\cite{note-program}

\section{K-edge RIXS formula in the limit of localized 1s core hole}\label{sec:rixs-formula}

\subsection{Derivation of the K-edge RIXS formula}\label{ssec:rixs-deriv}

The Kramers-Heisenberg formula~\cite{Ament11,Kramers25} is the starting point for the derivation of our
K-edge RIXS formula:
\begin{eqnarray}
    I(\omega,{\bf k},{\bf k}',\pmb{\epsilon},\pmb{\epsilon}') \propto \sum_{f}
    && \left| \sum_{n}
        \frac{ \langle f| \mathcal{D'}^{\dag}|n \rangle \langle n| \mathcal{D}|g \rangle }
        {E_g+\hbar \omega_{\bf k}-E_n + i \Gamma_n}
        \right|^2 \ \times \nonumber \\
    && \delta(E_g-E_f+\hbar\omega),
    \label{eq:KH}
\end{eqnarray}
where $|f\rangle$, $|n\rangle$, and $|g\rangle$ represent the final, intermediate and initial states,
$E_f$, $E_n$ and $E_g$ their energies, $\Gamma_n$ the inverse of the intermediate state lifetime,
$\hbar\omega_{\bf k}$ the energy of incoming x-ray with wavevector ${\bf k}$, and $\hbar\omega$ the x-ray
energy loss. $\mathcal{D'}^{\dag}$ and $\mathcal{D}$ are the electric multipole operators, which include
the incoming and outgoing x-ray wavevectors and polarization vectors, $({\bf k},\pmb{\epsilon})$ and
$({\bf k}',\pmb{\epsilon}')$.

In general, the $1s$ core-hole component of the intermediate eigenstates $|n\rangle$ can be chosen as
a delocalized momentum eigenstate.~\cite{Semba08} In the limit of the $1s$ electron hopping amplitude
approaching zero, the intermediate energy eigenstates with different core hole momenta become degenerate,
and the appropriate linear combinations can be made to form intermediate energy eigenstates with a $1s$ core
hole completely localized at a chosen site.~\cite{Ahn09,Davis79,Feldkamp80} Therefore, the state $|n\rangle$
can be chosen as $|n^{{\bf R}+{\bf d}} \rangle$, the intermediate energy eigenstate with the core hole at
a site ${\bf R}+{\bf d}$, where ${\bf R}$ and ${\bf d}$ represent the lattice point and the relative
position of core-hole site within the unit cell, respectively. The summation over intermediate states
$\sum_n$ can then be written as three summations, $\sum_{\bf R} \sum_{\bf d} \sum_{n^{{\bf R}+{\bf d}}}$.

\begin{figure}
    \centering
    \includegraphics[width=\hsize,clip]{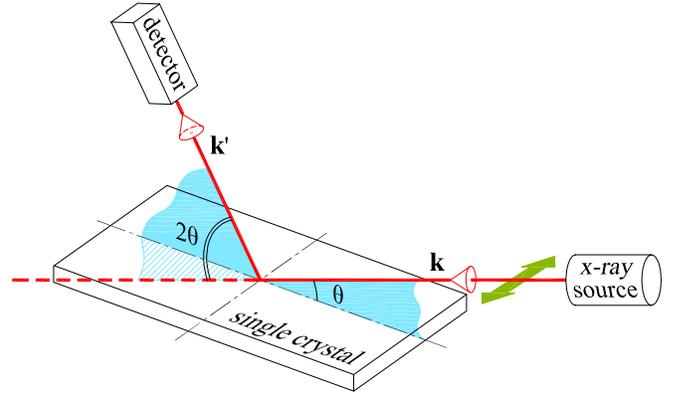}
    \caption{(Color online)
    Schematic drawing of the geometry for the K-edge RIXS experiment reported in this paper.
    The polarization direction is orthogonal to the scattering plane and is depicted with a green arrow.
    } \label{fig:beam-xray}
\end{figure}

We take the dipole approximation~\cite{Ament11} for the electric multipole operators $\mathcal{D'}^{\dag}$
and $\mathcal{D}$. By analyzing how the phases of the intermediate and final eigenstates change following a
translation by the lattice vector ${\bf R}$, we find that the sum over ${\bf R}$ gives rise to conservation of
crystal momentum. Under the appropriate experimental conditions, such as for the experiments reported in this
paper in which the scattering plane is fixed with respect to the crystal and the incoming x-ray polarization
vectors remain perpendicular to the scattering plane as shown in Fig.~\ref{fig:beam-xray},
the polarization effect in the K-edge RIXS is a constant
factor. We can then effectively remove the $4p$ creation and annihilation operators and replace the dipole
operators by the core-hole creation and annihilation operators. This results in the following expression,
\begin{eqnarray}
    && I(\omega,{\bf k},{\bf k}') \propto \nonumber \\
    && \sum_{\bf K} \sum_{f}
    \left|
        \sum_{\bf d} \sum_{n^{\bf d}}
        \frac{ e^{-i ({\bf k}'-{\bf k}) \cdot {\bf d}}
            \langle f| \underline{s}_{\bf d} |n^{\bf d} \rangle
            \langle n^{\bf d}| \underline{s}_{\bf d}^{\dagger} |g \rangle }
        {E_g+\hbar \omega_{\bf k}-E_{n^{\bf d}} + i \Gamma_{n^{\bf d}}}
    \right|^2 \times \nonumber \\
    && \delta(E_g-E_f+\hbar\omega) \ \delta({\bf k}_f+{\bf k}'-{\bf k}+{\bf K}),
    \label{eq:KH-IRIXS}
\end{eqnarray}
where $\underline{s}_{\bf d}^{\dagger}$ is the creation operator of the $1s$ core-hole at site ${\bf d}$,
${\bf K}$ represents a reciprocal lattice vector,
and ${\bf k}_f$ denotes the net momentum of the final state. 
Details of the derivation of the above formula are presented
in Appendix~\ref{sec:app-rixs}.

We make further approximations to simplify the numerical calculation of RIXS spectrum. First, we replace
the sum $\sum_{n^{\bf d}}$ by a single term with $n^{\bf d}=n^{\bf d}_{\rm low}$, that is, the lowest
energy eigenstate with the core hole at a site ${\bf d}$. This is justified for two reasons:~\cite{Ahn09}
First, $\langle n^{\bf d}| \underline{s}_{\bf d}^{\dagger} |g \rangle$ is largest for $n^{\bf d}_{\rm low}$,
the well-screened state. Second, the incoming x-ray energy is tuned to the absorption edge, which makes
the lowest energy intermediate state the most probable, while higher energy intermediate states, in
particular, the unscreened state,~\cite{Ahn09} are less likely to be excited by the incoming x-rays in
a K-edge RIXS process. The lowest energy intermediate state $|n^{\bf d}_{\rm low}\rangle$ is dominated by
the single-pair electron-hole excitations, especially in insulators, because of the higher energies necessary for
multiple-pair electron-hole excitations.~\cite{Ahn09} We therefore consider single-pair final states
$\langle l_e {\bf k}_e l_h {\bf k}_h \sigma |$ with an electron with wavevector ${\bf k}_e$, band index
$l_e$, and energy $\varepsilon_{l_e{\bf k}_e}$ and a hole with wavevector ${\bf k}_h$, band index $l_h$, and
energy $\varepsilon_{l_h{\bf k}_h}$, both with spin $\sigma$. Finally, if the resonant energies
$E_{n^{\bf d}}-E_g$ and core-hole lifetime broadenings $\Gamma_{n^{\bf d}}$ are similar for different core-hole
sites within the unit cell, then we can neglect the denominator in Eq.~(\ref{eq:KH-IRIXS}) for fixed incoming
x-ray energy, because it becomes a constant factor in the overall K-edge RIXS spectrum. These approximations
lead to the following formula, which we use for the numerical calculation of the K-edge RIXS spectrum.
\begin{eqnarray}
    && I(\omega,{\bf Q}) \propto \nonumber \\
    && \sum_{\bf K} \sum_{l_e {\bf k}_e l_h {\bf k}_h \sigma}
    \left|
        \sum_{\bf d}
         e^{-i {\bf Q} \cdot {\bf d}}
        \langle l_e {\bf k}_e l_h {\bf k}_h \sigma| \underline{s}_{\bf d} |n^{\bf d}_{\rm low} \rangle
        \langle n^{\bf d}_{\rm low}| \underline{s}_{\bf d}^{\dagger} |g \rangle
    \right|^2 \!\!\!\! \times \nonumber \\
    && \delta(\varepsilon_{l_h {\bf k}_h} - \varepsilon_{l_e {\bf k}_e} + \hbar\omega) \
    \delta({\bf k}_e-{\bf k}_h+{\bf Q}+{\bf K}),
    \label{eq:IRIXS}
\end{eqnarray}
where ${\bf Q}={\bf k}'-{\bf k}$. This formula relates the K-edge RIXS spectrum to the response of
the system to a localized charge. The reasonable approximations we have taken significantly reduce
the time for numerical calculations, and therefore, this formula can be used with the density functional
approach as well as the tight-binding approach. Details on how we numerically calculate $I(\omega,\bf{Q})$
with Eq.~(\ref{eq:IRIXS}) are presented in Appendix~\ref{sec:app-H-itok}.

\subsection{Periodicity of K-edge RIXS in reciprocal space}\label{ssec:rixs-kspace}

Understanding the periodicity of the K-edge RIXS spectrum is useful, for example, in determining where to probe
in reciprocal space. Also, as we will show below, there is useful information in the momentum dependence.
First, it should be noted that periodicity in reciprocal space is not inherent in inelastic x-ray scattering.
For example, off resonance, an increase in transferred momentum changes the transition matrix elements, since
higher order terms in the multipole expansion of the vector potential ${\bf A}$ are no longer
negligible.~\cite{Ament11} However,  on resonance, the matrix elements are all in the dipole limit and should
not depend on the momenta of the incoming and outgoing photons. The only relevant momentum is then the crystal
momentum, and the K-edge RIXS cross section follows the symmetry of the Brillouin zone. This was noted
experimentally by Kim {\it et al.}~\cite{Kim07} in their study of high-T$_c$ cuprates. However, these materials
have only one transition metal site in the unit cell. Such periodicity may not be generally applicable to
crystals with multiple core-hole sites within a unit cell, such as charge-orbital ordered manganites.

We look into the formula in Eq.~(\ref{eq:IRIXS}) to learn about the periodicity of K-edge RIXS spectrum.
For solid systems with one core-hole site per unit cell, we can choose ${\bf d}=0$ and simplify
Eq.~(\ref{eq:IRIXS}) by omitting a constant factor
$\langle n^{\bf d}_{\rm low}| \underline{s}_{\bf d}^{\dagger} |g \rangle$ to obtain,
\begin{eqnarray}
    I(\omega,{\bf Q})\propto && \sum_{\bf K} \sum_{l_e {\bf k}_e l_h {\bf k}_h \sigma}
    \left|
        \langle l_e {\bf k}_e l_h {\bf k}_h \sigma | \underline{s}_{{\bf d}=0} |n^{{\bf d}=0}_{\rm low} \rangle
    \right|^2 \ \times \nonumber \\
    && \delta(\varepsilon_{l_h {\bf k}_h} - \varepsilon_{l_e {\bf k}_e} + \hbar\omega) \
    \delta({\bf k}_e-{\bf k}_h+{\bf Q}+{\bf K}),
    \label{eq:IRIXS-2}
\end{eqnarray}
which makes the RIXS calculations even simpler for materials with one core-hole site per unit cell.
If the x-ray wavevector change ${\bf Q}$ is altered by a reciprocal lattice vector ${\bf K}'$, i.e.,
${\bf Q}'={\bf Q}+{\bf K}'$, then the RIXS intensity will be unchanged, since ${\bf K}''={\bf K}'+{\bf K}$
in the second $\delta$-function is also a reciprocal lattice vector. This is consistent with the experimental
result for La$_2$CuO$_4$ (Ref.~\onlinecite{Kim07}).

On the other hand, if a solid has multiple core-hole sites per unit cell
due to the ordering of spin, charge, orbital, or local lattice distortions,
then the following argument shows that the symmetry of the K-edge
RIXS spectrum is with respect to the lattice without ordering, rather than the actual lattice. 
We represent
the lattice without ordering by ${\bf R}_{\rm core}$, which includes 
the actual lattice ${\bf R}$ as well as ${\bf d}$. Then ${\bf K}_{\rm core}$,
the reciprocal lattice vector of ${\bf R}_{\rm core}$, satisfies the condition of
$e^{i {\bf K}_{\rm core} \cdot {\bf d}} = 1$, which results in the symmetry of K-edge RIXS
spectrum in Eq.~(\ref{eq:IRIXS}), that is, $I(\omega,{\bf Q}+{\bf K}_{\rm core}) = I(\omega,{\bf Q})$.
We shall see this explicitly in Sec.~\ref{ssec:rixs-periodic} for La$_{0.5}$Sr$_{1.5}$MnO$_4$.


\section{Experimental results for
L\lowercase{a}$_{0.5}$S\lowercase{r}$_{1.5}$M\lowercase{n}O$_4$}\label{sec:exper}

\begin{figure}
    \centering
    \includegraphics[width=0.95\hsize,clip]{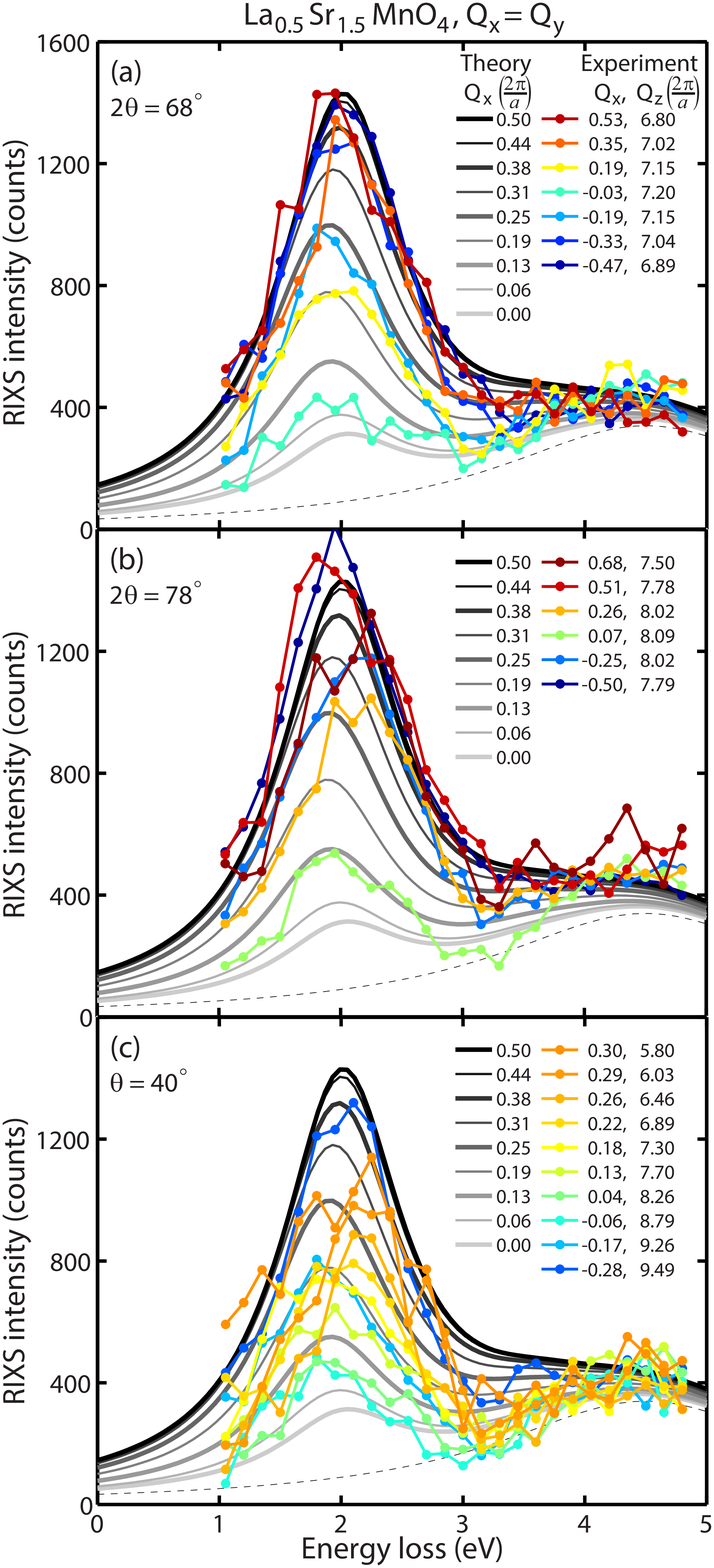}
    \caption{(Color online)
    Experimental and theoretical K-edge RIXS intensities for La$_{0.5}$Sr$_{1.5}$MnO$_4$ with CE-type
    charge-orbital-spin ordering. The symbols in panels (a), (b) and (c) represent the experimental data,
    taken at $2\theta=68^{\circ}$, $2\theta=78^{\circ}$, and $\theta=40^{\circ}$ respectively, and
    $Q_x = Q_y = H_{\rm ex}\frac{2\pi}{a}$ and $Q_z = L_{\rm ex}\frac{2\pi}{a}$. The elastic peaks
    are subtracted from data.~\cite{Liu13} Dashed line represents a momentum-independent peak, presumably
    from a O-Mn transition. Solid curves in gray scale represent the RIXS intensity calculated at
    $Q_x = Q_y = H_{\rm th}\frac{2\pi}{a}$ for the Mn-Mn transition for the parameter set with
    $t_0$ = 0.9~eV, added to the dashed line, as discussed in Sec.~V~D.
    } \label{fig:exp-vs-theor}
\end{figure}

Mn K-edge RIXS from La$_{0.5}$Sr$_{1.5}$MnO$_4$ was measured at the Advanced Photon Source on beamlines 30-ID
and 9-ID at temperature $T$ = 20 K, well below CE-type magnetic, charge, orbital, and structural ordering
temperatures. The instrumental energy resolution was of about 270~meV (FWHM). As shown in
Fig.~\ref{fig:beam-xray}, a single crystal grown in traveling solvent floating zone method is aligned so that
when the x-ray wavevector transfer ${\bf Q}$ is in the scattering plane, it has $Q_x=Q_y$ with the $x$ and $y$
axes along the Mn-O bond direction and the $z$ axis perpendicular to the MnO$_2$ plane. The scattering plane
was fixed with respect to the crystal and the polarization of the incoming and outgoing x-rays was
perpendicular to the scattering plane, so that the polarization factor is a constant factor in the RIXS
formula, as assumed in the derivation of Eq.~(\ref{eq:IRIXS}). Data taken either at a fixed sample angle,
$\theta$, or a fixed detector angle, 2$\theta$, are shown as connected dots in Fig.~\ref{fig:exp-vs-theor}. The elastic peak
has been subtracted from the data.~\cite{Liu13} The main focus in this paper is the intensity variation of
the 2~eV peak, which is known to arise from transitions between Mn $3d$ $e_g$ bands from optical
measurements.~\cite{Ishikawa99,Jung00} The intensity of the 2~eV peak increases rapidly from $Q_x = Q_y = 0$
to $Q_x = Q_y = \pm\frac{\pi}{a}$, where $a$ represent the average Mn-Mn distance within the MnO$_2$ plane,
but is almost independent of $Q_z$. While the latter supports the two-dimensional character of the $e_g$
electrons confined within each MnO$_2$ layer, the former cannot be explained by dynamic structure factor
$S({\bf q},\omega)$. This, combined with the fact that there are eight core-hole sites per two-dimensional
unit cell, makes the experimental results for La$_{0.5}$Sr$_{1.5}$MnO$_4$ an ideal case to test the validity
of our theory.

\section{Tight binding Hartree-Fock Hamiltonian and core-hole potential for \lowercase{$e_g$} electrons in
L\lowercase{a}$_{0.5}$S\lowercase{r}$_{1.5}$M\lowercase{n}O$_4$}\label{sec:tight-bind}

\begin{figure}
    \centering
    \includegraphics[width=\hsize,clip]{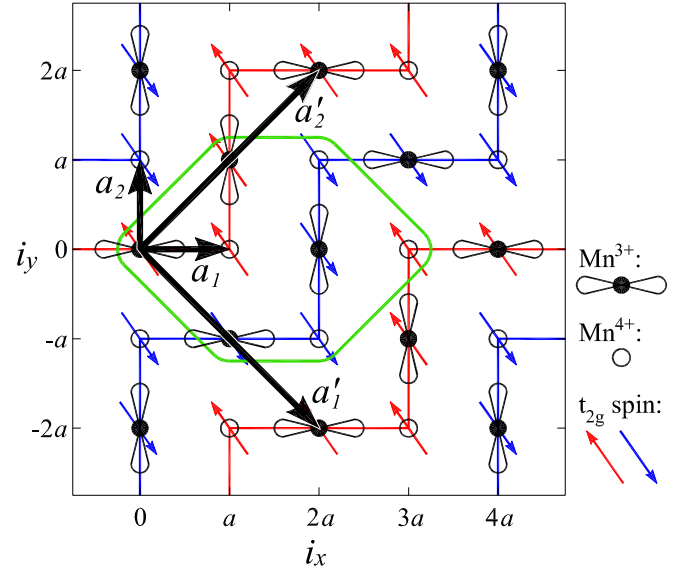}
    \caption{(Color online)
    In-plane structural layout of Mn ions for La$_{0.5}$Sr$_{1.5}$MnO$_4$ in CE-type ordering. Red and blue
    arrows represent the $t_{2g}$ spin alignment. ${\bf a}'_1$ and ${\bf a}'_2$ represent the primitive
    lattice vectors for CE-phase. ${\bf a}_1$ and ${\bf a}_2$ are primitive vectors for lattice without
    charge, orbital, and spin ordering. The rounded hexagon encloses the Mn ions in the basis.
    } \label{fig:ce-order}
\end{figure}

La$_{0.5}$Sr$_{1.5}$MnO$_4$ has a layered two-dimensional perovskite structure with negligible hopping of
the Mn $3d$ $e_g$ electrons between the MnO$_2$ layers. This is consistent with the experimental observation
of K-edge RIXS spectrum being independent of the momentum transfer perpendicular to MnO$_2$ layers.
We therefore consider a Hamiltonian for a single MnO$_2$ layer. La$_{0.5}$Sr$_{1.5}$MnO$_4$ undergoes
a structural and orbital ordering transition at 230 K, and a CE-type magnetic ordering transition at 110~K,
schematically shown in Fig.~\ref{fig:ce-order} for the MnO$_2$ layer. In this figure, ``Mn$^{3+}$" and
``Mn$^{4+}$" are used to indicate the two sites not related by symmetry, rather than controversial charge
ordering.~\cite{Herrero04,Brey04,Coey04} The strong Hund's coupling between the $e_g$ electron spin and
the $t_{2g}$ electron spin confines most of the $e_g$ electron hopping along the zigzag chain. The distortion
of the oxygen octahedron surrounding the Mn ions splits the $e_g$ energy levels through the Jahn-Teller
electron-lattice coupling.

Our tight-binding Hamiltonian considers the \textit{effective} Mn $3d$ $e_g$ levels only, because the RIXS
peak at around 2~eV is due to transitions between the bands from these levels.~\cite{Ishikawa99,Jung00} 
We note that these effective
Mn $3d$ $e_g$ levels are in fact linear combinations of \textit{atomic} Mn $3d$ $e_g$ levels and {\it atomic}
O $2p$ levels. Appendix~\ref{sec:app-eh-number} discusses this aspect in more detail, in particular,
in relation to the electron numbers on the Mn ions.

Within this effective model, we define $d^{\dagger}_{{\bf i}\xi\sigma}$ as the creation operator of
the $e_g$ electron with the spin state $\sigma$ = $\uparrow$, $\downarrow$ and orbital state $\xi = 1$ for
$(3z^2-r^2)/2$ and $\xi = 2$ for $\sqrt{3}(x^2-y^2)/2$ at the Mn site ${\bf i}=(m_x a,m_y a)$, where
$m_x$ and $m_y$ are integers,
as shown in Fig.~\ref{fig:ce-order}. The electron hopping term~\cite{Ahn00} is
\begin{equation}
    \hat{H}_{\rm hopping} = -\frac{1}{2}\sum_{{\bf i},\pmb{\delta},\xi,\xi',\sigma} \!\!\!
        t_{\pmb{\delta}}^{\xi\xi'}
        \left( d_{{\bf i}\xi\sigma}^{\dagger } d_{{\bf i}+\pmb{\delta},\xi',\sigma} +
        d_{{\bf i}+\pmb{\delta},\xi',\sigma}^{\dagger} d_{{\bf i}\xi\sigma} \right).
\end{equation}
The vector $\pmb{\delta} = \pm a\hat{x},\pm a\hat{y}$ represents the nearest neighbor sites of
a Mn ion. The hopping matrices within the MnO$_2$ plane are
\begin{eqnarray}
    t_{a\hat{x}} &=&t_{-a\hat{x}}=t_{0}\left(
        \begin{array}{cc}
        1/4 & -\sqrt{3}/4 \\
        -\sqrt{3}/4 & 3/4
        \end{array}
    \right) , \\
    t_{a\hat{y}} &=&t_{-a\hat{y}}=t_{0}\left(
        \begin{array}{cc}
        1/4 & \sqrt{3}/4 \\
        \sqrt{3}/4 & 3/4
        \end{array}
    \right),
\end{eqnarray}
reflecting the symmetry of the $e_g$ orbitals. The parameter $t_0$ represents the effective hopping
constant between the two $(3x^2-r^2)/2$ orbitals along the $x$-direction.

\begin{figure}
    \centering
    \includegraphics[width=\hsize,clip]{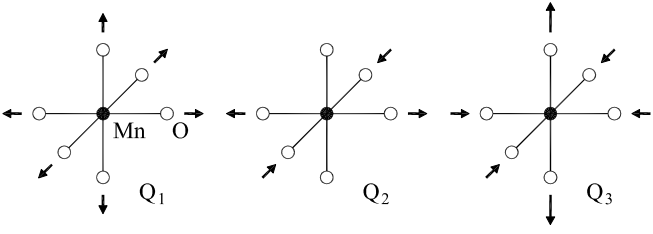}
    \caption{
    Distortion modes $Q_1$, $Q_2$, and $Q_3$ of oxygen octahedron around Mn with arrows indicating
    directions of displacement.
    } \label{fig:Q1Q2Q3}
\end{figure}

The distortion of oxygen octahedron around a Mn ion at site ${\bf i}$ is parameterized as follows.
$u_{\bf i}^{\zeta }$ ($\zeta $ = $x,y$) represents the $\hat{\zeta}$ directional displacement of
an oxygen ion located between Mn ions at ${\bf i}$ and ${\bf i}+a\hat{\zeta }$ from the position
for the ideal undistorted square MnO$_2$ lattice with the average in-plane Mn-O bond distance.
The $u_{\bf i}^{+z}$ and $u_{\bf i}^{-z}$ represent the $z$ direction displacements of the oxygen ions
above and below the Mn ion at site ${\bf i}$ from the location of the average in-plane Mn-O
bond distance. The parameters, $Q_{1{\bf i}}$, $Q_{2{\bf i}}$, and $Q_{3{\bf i}}$, represent
the distortion modes of the oxygen octahedron, shown in Fig.~\ref{fig:Q1Q2Q3} and defined as follows.
\begin{eqnarray}
    Q_{1{\bf i}} &=& \frac{ u^x_{\bf i}-u^x_{{\bf i}-a\hat{x}}+u^y_{\bf i}-u^y_{{\bf i}-a\hat{y}}
        +u^{+z}_{\bf i}-u^{-z}_{\bf i} }{\sqrt{3}}, \\
    Q_{2{\bf i}} &=& \frac{ u^x_{\bf i}-u^x_{{\bf i}-a\hat{x}}-u^y_{\bf i}+u^y_{{\bf i}-a\hat{y}} }{\sqrt{2}}, \\
    Q_{3{\bf i}} &=& \frac{ 2 u^z_{\bf i}- 2 u^{-z}_{\bf i}
        -u^x_{\bf i}+u^x_{{\bf i}-a\hat{x}}
        -u^y_{\bf i}+u^y_{{\bf i}-a\hat{y}} }{\sqrt{6}}.
\end{eqnarray}

The Mn-O bond distances estimated from structural refinement of high-resolution synchrotron x-ray powder
diffraction data for La$_{0.5}$Sr$_{1.5}$MnO$_4$ in Ref.~\onlinecite{Herrero11} indicate $Q_1=0.0531$~{\AA},
$Q_2=\pm 0.1089$~{\AA}, and $Q_3=0.0955$~{\AA} around the ``Mn$^{3+}$" sites and $Q_1=-0.0531$~{\AA},
$Q_2=0$, and $Q_3=0.1192$~{\AA} around the ``Mn$^{4+}$" sites.

The $Q_2$ and $Q_3$ distortions break the cubic symmetry of the oxygen octahedra and interact with the $e_g$
orbital state through the following Jahn-Teller Hamiltonian term,~\cite{note-JT}
\begin{equation}
    \hat{H}_{\rm JT} = -\lambda \sum_{{\bf i},\sigma}\left(
        \begin{array}{c}
        d_{{\bf i}1\sigma}^{\dagger} \\
        d_{{\bf i}2\sigma}^{\dagger}
        \end{array}
    \right)^{T} \left(
        \begin{array}{cc}
         Q_{3{\bf i}} & -Q_{2{\bf i}} \\
        -Q_{2{\bf i}} & -Q_{3{\bf i}}
        \end{array}
    \right) \left(
        \begin{array}{c}
        d_{{\bf i}1\sigma} \\
        d_{{\bf i}2\sigma}
        \end{array}
    \right),
\end{equation}
where $\lambda$ corresponds to the Jahn-Teller coupling constant. The isotropic $Q_1$ distortion interacts
with the total $e_g$ electron number through the following ``breathing" electron-lattice Hamiltonian
term,~\cite{Millis96}
\begin{equation}
    \hat{H}_{\rm br} = -\beta \lambda \sum_{{\bf i},\sigma }\left(
        \begin{array}{c}
        d_{{\bf i}1\sigma}^{\dagger} \\
        d_{{\bf i}2\sigma}^{\dagger}
        \end{array}
    \right) ^{T}\left(
        \begin{array}{cc}
        Q_{1{\bf i}} & 0 \\
        0 & Q_{1{\bf i}}
        \end{array}
    \right) \left(
        \begin{array}{c}
        d_{{\bf i}1\sigma} \\
        d_{{\bf i}2\sigma}
        \end{array}
    \right),
\end{equation}
where $\beta$ represents the ratio between the strengths of the breathing and the Jahn-Teller coupling.

We also include the Hund's coupling of the $e_g$ electron spin state to the classical $t_{2g}$ spin direction,
\begin{equation}
    \hat{H}_{\rm Hund} = -J_H \sum_{{\bf i},\xi,\sigma',\sigma''}
        {\bf S}_{t_{2g}{\bf i}}\cdot
        d_{{\bf i}\xi\sigma'}^{\dagger } \  \pmb{\tau}_{\sigma'\sigma''} \ d_{{\bf i}\xi\sigma''},
\end{equation}
where $J_H$ represents the Hund's coupling constant, ${\bf S}_{t_{2g}{\bf i}}$ the $t_{2g}$ spin vector
ordered in CE-type structure and $\pmb{\tau}$ is the Pauli matrix vector.

As in Ref.~\onlinecite{Ahn00}, we also include the $3d$-$3d$ on-site Coulomb interaction,
\begin{equation}
    \hat{H}_{\rm dd} = U \sum_{\bf i}\sum_{(\eta,\sigma)\neq(\eta',\sigma')}
        \hat{n}_{{\bf i}\eta\sigma}\hat{n}_{{\bf i}\eta'\sigma'},
\end{equation}
where $\hat{n}_{{\bf i}\eta\sigma} = d^{\dag}_{{\bf i}\eta\sigma} d_{{\bf i}\eta\sigma}$ is the number
operator and $U$ represents the size of the $3d$-$3d$ Coulomb interaction. The index $\eta=-,+$ represents
the local orbital eigenstates of $\hat{H}_{\rm JT}$ with lower and higher energies, respectively, chosen
for the following Hartree-Fock approximation:
\begin{eqnarray}
    \hat{H}_{\rm dd}^{\rm HF} = \sum_{\bf i} \bigg(
         U_{{\bf i}+\uparrow} d_{{\bf i}+\uparrow}^{\dagger} d_{{\bf i}+\uparrow}
        +U_{{\bf i}-\uparrow} d_{{\bf i}-\uparrow}^{\dagger} d_{{\bf i}-\uparrow} \nonumber \\
        +U_{{\bf i}+\downarrow} d_{{\bf i}+\downarrow}^{\dagger} d_{{\bf i}+\downarrow}
        +U_{{\bf i}-\downarrow} d_{{\bf i}-\downarrow}^{\dagger} d_{{\bf i}-\downarrow} \bigg),
\end{eqnarray}
where $U_{{\bf i}+\uparrow} = U (\langle\hat{n}_{{\bf i}-\uparrow}\rangle +
\langle\hat{n}_{{\bf i}+\downarrow}\rangle + \langle\hat{n}_{{\bf i}-\downarrow}\rangle)$, etc.~\cite{Ahn00}

The total Hamiltonian for Mn $3d$ $e_g$ electrons for calculations of K-edge RIXS initial and final states is
then the sum of the terms described so far,
\begin{equation}\label{eq:H3d}
    \hat{H}_{\rm d} = \hat{H}_{\rm hopping} + \hat{H}_{\rm JT} + \hat{H}_{\rm br}
        + \hat{H}_{\rm Hund} +\hat{H}_{\rm dd}^{\rm HF}.
\end{equation}
The CE type ordering of the $t_{2g}$ spins and the lattice distortions associated with charge and
orbital ordering give rise to the primitive lattice vectors ${\bf a}'_1$ and ${\bf a}'_2$ shown in
Fig.~\ref{fig:ce-order}. The primitive reciprocal lattice vectors are
${\bf b}'_1=(\frac{\pi}{2a},-\frac{\pi}{2a})$ and ${\bf b}'_2=(\frac{\pi}{2a},\frac{\pi}{2a})$, and
the first Brillouin zone is $\Omega_{\rm 1BZ}=\{{\bf k}| -\frac{\pi}{2a}<k_x+k_y\le\frac{\pi}{2a}$,
$-\frac{\pi}{2a}<k_x-k_y\le\frac{\pi}{2a}\}$.

In the intermediate state, we must also account for the presence of the core hole. The $1s$-$3d$ on-site
Coulomb interaction is generally expressed as
\begin{equation}
    \hat{H}_{{\rm sd}} = -U_{\rm core} \sum_{{\bf i},\xi,\sigma,\sigma'}
        d^{\dagger}_{{\bf i}\xi\sigma} \ d_{{\bf i}\xi\sigma} \
        \underline{s}^{\dagger}_{{\bf i}\sigma'} \ \underline{s}_{{\bf i}\sigma'},
\end{equation}
where $U_{\rm core}$ represents the size of the $1s$-$3d$ Coulomb interaction, and
$\underline{s}^{\dagger}_{{\bf i}\sigma'}$ is the creation operator for a $1s$ core hole with spin $\sigma'$
at site ${\bf i}$. As discussed in Sec.~\ref{ssec:rixs-deriv}, in the limit of vanishing $1s$ electron hopping
amplitude, the K-edge RIXS intermediate energy eigenstates can be chosen as states with a single completely
localized $1s$ core hole, which can be found from
\begin{equation}\label{eq:Htot}
    \hat{H}_{{\rm total},{\bf i}_c} = \hat{H}_{\rm d} + \hat{H}_{{\rm sd},{\bf i}_c},
\end{equation}
with
\begin{equation}
    \hat{H}_{{\rm sd}, {\bf i}_c} = -U_{\rm core}\sum_{\xi,\sigma}
        d^{\dagger}_{{\bf i}_c\xi\sigma} \ d_{{\bf i}_c\xi\sigma},
\end{equation}
${\bf i}_c$ representing the $1s$ core-hole site, and
$\underline{s}^{\dagger}_{{\bf i}_c\sigma'} \ \underline{s}_{{\bf i}_c\sigma'}=1$ being used. To calculate
the K-edge RIXS spectrum, we need to represent the eigenstates of $\hat{H}_{{\rm total},{\bf i}_c}$, as
a linear combination of the eigenstates of $\hat{H}_{\rm d}$, as described in detail in
Appendix~\ref{sec:app-H-itok}. The expression of the Hamiltonian $\hat{H}_{\rm d}$ and
$\hat{H}_{{\rm total}, {\bf i}_c}$ in reciprocal space for La$_{0.5}$Sr$_{1.5}$MnO$_4$ is presented in
Appendix~\ref{sec:app-hamilt}.

Each Hamiltonian term has one parameter. Some of the parameter values are chosen by modifying
corresponding values for LaMnO$_3$ found in Ref.~\onlinecite{Ahn00}. The chosen parameter values
are $t_0$ = 0.9~eV, $\lambda$ = 7.41~eV/{\AA}, $\beta$ = 1.5, $J_H |{\bf S}_{t_{2g},{\bf i}}|$ = 2.2~eV,
$U$ = 3.5~eV, and $U_{\rm core}$ = 4.0~eV. In addition, we vary $t_0$ and $\lambda$ while maintaining
the gap size around 2~eV to examine how the RIXS spectrum depends on the $e_g$ electron hopping
amplitude.

\section{Results from theory and comparison with experiments}\label{sec:results}

\subsection{Electronic density of states in the absence and in the presence of the core hole}

\begin{figure}
    \centering
    \includegraphics[width=\hsize,clip]{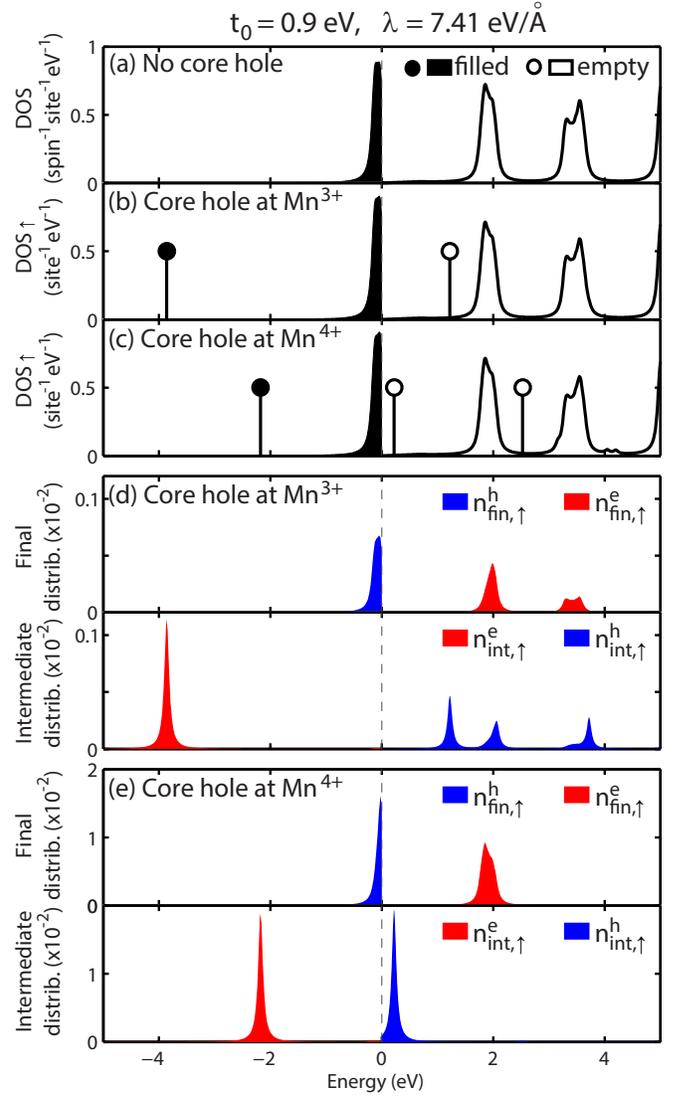}
    \caption{(Color online)
    Electronic density of states (DOS) per site, for $t_0=0.9$~eV and $\lambda=7.41$~eV/{\AA}
    (a) without a core hole,
    (b) with a core hole at a Mn$^{3+}$ site and
    (c) with a core hole at a Mn$^{4+}$ site.
    The Lorentzian broadening of 2$\Gamma$ = 0.1~eV is used to make the DOS curve smooth. Vertical lines
    with a circle on top represent bound states.
    (d) and (e): Final (upper panel) and intermediate (lower panel) electron-hole distribution
    with a core hole at a (d) Mn$^{3+}$ and (e) Mn$^{4+}$ site.
    } \label{fig:dos-nenh-t0.9}
\end{figure}

We first present our results on energy eigenstates and eigenvalues of the Hamiltonians for a 16$\times$16
Mn site cluster with periodic boundary conditions. The calculated density of states (DOS) is shown in
Fig.~\ref{fig:dos-nenh-t0.9}(a) in the absence of a core hole. The occupied band mostly consists of
the lower Jahn-Teller $e_g$ levels with spin parallel to the $t_{2g}$ spins at Mn$^{3+}$ sites, whereas
the lowest empty band mostly consists of similar $e_g$ levels at Mn$^{4+}$ sites. The excitation
between these two bands is responsible for the 2~eV RIXS peak, which is the focus of our comparison
with experiment data. Due to spin degeneracy in CE-type antiferromagnetic ordering, the electronic
DOS $D_{\downarrow}(\varepsilon)$ for spin $\downarrow$, is identical to that for spin $\uparrow$,
$D_{\uparrow}(\varepsilon)$.
\begin{figure}
    \centering
    \includegraphics[width=\hsize,clip]{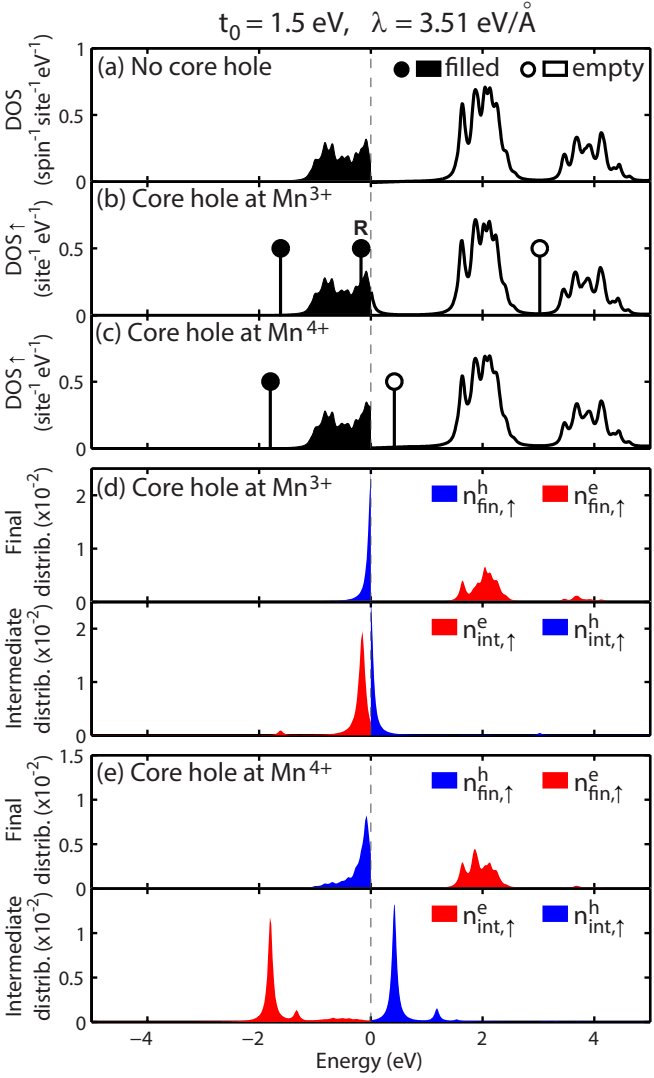}
    \caption{(Color online)
    Similar figures as Fig.~\ref{fig:dos-nenh-t0.9} for $t_0=1.5$~eV and $\lambda=3.51$~eV/{\AA}.
    ``R" in (b) represents a resonant state, rather than a bound state. Note that a single state at
    the top of the lower band in (b) is empty.
    } \label{fig:dos-nenh-t1.5}
\end{figure}

We next analyze the Hamiltonian $\hat{H}_{{\rm total},{\bf i}_c}$ in the presence of the core hole at
site ${\bf i}_c$. The $t_{2g}$ spin direction at ${\bf i}_c$ breaks the spin degeneracy in the DOS.
The DOS $D_{\uparrow}^{{\bf i}_c=(0,0)}(\varepsilon)$ is displayed in Fig.~\ref{fig:dos-nenh-t0.9}(b),
for the core hole site ${\bf i}_c = (0,0)$ in Fig.~\ref{fig:ce-order}, which is a Mn$^{3+}$ site with
$t_{2g}$ spin $\uparrow$. The core-hole potential pulls out bound states from the band continuum,~\cite{Ahn09}
identified by vertical lines with circles on top in Fig.~\ref{fig:dos-nenh-t0.9}(b). The lowest bound
state is at about $-4$~eV, that is, $U_{\rm core}$ below the lowest band with Mn$^{3+}$ character. The second
bound state is within the gap. The DOS for the band continuum is almost unchanged, except that the number
of states within each band below and above the gap is reduced by one because of the bound states pulled
out.~\cite{Ahn09} By filling the states from the lowest energy with the same number of electrons in
the intermediate states as in the ground state, as shown in Fig.~\ref{fig:dos-nenh-t0.9}(b), we obtain
the lowest energy intermediate state, $\underline{s}|n_{\rm low}^{{\bf i}_c}\rangle$. Therefore, the bound
state below the lowest band is occupied and the bound state within the gap is empty in the intermediate state.
$D_{\downarrow}^{{\bf i}_c=(0,0)}(\varepsilon)$ is almost identical to the DOS without a core hole in
Fig.~\ref{fig:dos-nenh-t0.9}(a), because the electrons with spin $\downarrow$ contribute very little to
the screening of the core hole due to the strong Hund's coupling.

The DOS $D_{\uparrow}^{{\bf i}_c=(a,0)}(\varepsilon)$, for the core hole at ${\bf i}_c = (a,0)$ in
Fig.~\ref{fig:ce-order}, which is a Mn$^{4+}$ site with $t_{2g}$ spin $\uparrow$, is shown in
Fig.~\ref{fig:dos-nenh-t0.9}(c) and has similar features. The lowest bound state is at around $-2$~eV,
that is, $U_{\rm core}$ below the band with the Mn$^{4+}$ site character, which is the lowest empty band.
Again, the lowest bound state is filled and the bound state within the gap is empty for
$\underline{s}|n_{\rm low}^{{\bf i}_c}\rangle$, as indicated in Fig.~\ref{fig:dos-nenh-t0.9}(c).

As a comparison, we carry out similar calculations for the parameter values of $t_0$ = 1.5~eV and
$\lambda$ = 3.51~eV/{\AA}, which keep the size of the gap approximately 2~eV, but result in a larger
band width. Due to the larger electron hopping, the gap has more of a hybridization gap character, and
the bands are wider, as shown in Fig.~\ref{fig:dos-nenh-t1.5}(a), which shows the DOS without
a core hole. Similarly to the $t_0$ = 0.9~eV case, Figs.~\ref{fig:dos-nenh-t1.5}(b) and
\ref{fig:dos-nenh-t1.5}(c) show $D_{\uparrow}^{{\bf i}_c}(\varepsilon)$, in the presence of a core
hole at ${\bf i}_c=(0,0)$ and ${\bf i}_c=(a,0)$, respectively. The bound states in
Fig.~\ref{fig:dos-nenh-t1.5}(c) are qualitatively similar to those for $t_0$ = 0.9~eV in
Fig.~\ref{fig:dos-nenh-t0.9}(c). However, qualitatively different behavior occurs for the core hole
at a Mn$^{3+}$ site, as shown in Fig.~\ref{fig:dos-nenh-t1.5}(b). In this case, the bound state
that would be in the gap for smaller $t_0$ resides in the occupied band and becomes a resonant
rather than bound state, as indicated by the vertical line with ``R" on top. Such a resonant state
hybridizes with delocalized states in the band, unlike bound states. With the bound state below
the lower band and this resonant state occupied, the top of the lower band is empty in the lowest
energy intermediate state $\underline{s}|n_{\rm low}^{{\bf i}_c}\rangle$, as indicated in
Fig.~\ref{fig:dos-nenh-t1.5}(b). This will have a significant consequence in the screening dynamics,
as discussed in following sections.

\subsection{Electron and hole excitations by the core hole represented along energy axis}\label{ssec:results-n}

Understanding the distributions of the electrons and holes that are excited by the core hole is
essential for the interpretation of RIXS spectrum. Excited electron and hole distributions with respect to
the energy for the RIXS \emph{final} state, $n_{{\rm fin},\sigma}^e(\varepsilon)$ with
$\varepsilon > \varepsilon_{F}$ and $n_{{\rm fin},\sigma}^h(\varepsilon)$ with
$\varepsilon < \varepsilon_{F}$, are defined as follows~\cite{Ahn09}
\begin{eqnarray}
    n_{{\rm fin},\sigma}^e(\varepsilon) &=& \!\!\!\!\!
        \sum_{l {\bf k}}^{\varepsilon_{l{\bf k}\sigma} > \varepsilon_{F}}
        \sum_{m=1}^{\frac{N_e}{2}}
        \left| \langle 0| b_{l{\bf k}\sigma} c_{m\sigma}^{\dagger} |0 \rangle \right|^2
        \delta(\varepsilon - \varepsilon_{l{\bf k}\sigma}), \label{eq:n-fin-e}\\
    n_{{\rm fin},\sigma}^h(\varepsilon) &=& \!\!\!\!\!
        \sum_{l {\bf k}}^{\varepsilon_{l{\bf k}\sigma} < \varepsilon_{F}} \!\!\!\!\!
        \sum_{m=\frac{N_e}{2}+1}^{\frac{N_e}{2}} \!\!\!\!\!
        \left| \langle 0| b_{l{\bf k}\sigma} c_{m\sigma}^{\dagger} |0 \rangle \right|^2
        \delta(\varepsilon - \varepsilon_{l{\bf k}\sigma}), \label{eq:n-fin-h}
\end{eqnarray}
where $\varepsilon_{F}$ represents the Fermi energy in the absence of the core hole,
$c_{m\sigma}^{\dagger}$ is the creation operator for the $m$-th lowest energy eigenstate
of $\hat{H}_{{\rm total},{\bf i}_c}$ with spin $\sigma$ and energy $\varepsilon_{m\sigma}$, and
$b_{l{\bf k}\sigma}^{\dagger}$ represents the creation operator for the energy eigenstate with wavevector
${\bf k}$ and energy $\varepsilon_{l{\bf k}\sigma}$ within the $l$-th lowest band of $\hat{H}_{\rm d}$, and
$N_e$ represents total electron number. These are the electrons and holes that are moved by the scattering
process. For example, $n_{\rm fin}^e(\varepsilon)$ corresponds to the projection of occupied intermediate
states to unoccupied initial states. The $\delta$-function makes this distribution represented with respect
to the energy without the core hole.

Similar electron and hole distributions with respect to the energy for the RIXS \emph{intermediate}
state,~\cite{Ahn09} $n_{{\rm int},\sigma}^e(\varepsilon)$ and $n_{{\rm int},\sigma}^h(\varepsilon)$, are
defined in the same way as Eqs.~(\ref{eq:n-fin-e}) and (\ref{eq:n-fin-h}), except that the energy
$\delta$-function is replaced by $\delta(\varepsilon - \varepsilon_{m\sigma})$. This difference also implies
that we have $n_{{\rm int},\sigma}^e(\varepsilon)$ for $\varepsilon < \varepsilon_{F}^{\rm int}$ and
$n_{{\rm int},\sigma}^h(\varepsilon)$ for $\varepsilon > \varepsilon_{F}^{\rm int}$, where
$\varepsilon_{F}^{\rm int}$ represents the Fermi energy in the presence of the core hole.

These excited electron and hole distributions are plotted in Figs.~\ref{fig:dos-nenh-t0.9}(d), \ref{fig:dos-nenh-t0.9}(e),
\ref{fig:dos-nenh-t1.5}(d), and \ref{fig:dos-nenh-t1.5}(e) for $\sigma$ = $\uparrow$.
Similar distributions for spin $\downarrow$ state are less than 10\% of those for spin $\uparrow$
state. The plots of $n^e_{\rm int,\uparrow}(\varepsilon)$ and $n^h_{\rm int,\uparrow}(\varepsilon)$
in the \emph{lower} panels of Figs.~\ref{fig:dos-nenh-t0.9}(d), \ref{fig:dos-nenh-t0.9}(e), and
\ref{fig:dos-nenh-t1.5}(e) show that the bound states in the intermediate state, marked in
Fig.~\ref{fig:dos-nenh-t0.9}(b), \ref{fig:dos-nenh-t0.9}(c), and \ref{fig:dos-nenh-t1.5}(c),
respectively, make the dominant contribution to the electron-hole excitations.

The plots of $n^h_{\rm fin,\uparrow}(\varepsilon)$ and $n^e_{\rm fin,\uparrow}(\varepsilon)$
in the \emph{upper} panels of Figs.~\ref{fig:dos-nenh-t0.9}(d), \ref{fig:dos-nenh-t0.9}(e) and
\ref{fig:dos-nenh-t1.5}(e), show that, while $n_{{\rm fin},\sigma}^e(\varepsilon)$ resembles the DOS
of the unoccupied band, $n_{{\rm fin},\sigma}^h(\varepsilon)$ shows a peak at the top of the occupied
band, in particular, in Fig.~\ref{fig:dos-nenh-t1.5}(e). This represents asymmetric screening dynamics
between electrons and holes.

For the exceptional case of $t_0$ = 1.5~eV and a core hole at Mn$^{3+}$ site, the comparison between
Fig.~\ref{fig:dos-nenh-t1.5}(b) and the lower panel in Fig.~\ref{fig:dos-nenh-t1.5}(d) reveals that
the occupied resonant state within the lower band marked by ``R" and the empty state at the top of the occupied
band in Fig.~\ref{fig:dos-nenh-t1.5}(b) make dominant contributions to $n^e_{\rm int,\uparrow}(\varepsilon)$
and $n^h_{\rm int,\uparrow}(\varepsilon)$. The resonant state is dominant for electron excitations because
this state is pulled from the initially unoccupied bands, whereas the lowest bound state comes mostly from
the initially occupied band. The delocalized state at the top of the occupied band predominantly contributes
to the hole excitation, because it is occupied in the initial state and empty in the intermediate state, which
results in hole excitation very close to the gap shown in the upper panel in Fig.~\ref{fig:dos-nenh-t1.5}(d).

We note that these results are consistent with the conclusions of Ref.~\onlinecite{Ahn09}, thus validating
the present study.

\subsection{Electron and hole excitations by the core hole represented in real space}

In this subsection, we present the distribution of electrons and holes excited by the core hole in real
space.

First, in the absence of the core hole, the electron number $\langle\hat{n}_{{\bf i}\eta\sigma}\rangle$
is calculated for each spin state $\sigma = \uparrow, \downarrow$ and orbital state $\eta = +, -$ at
each site ${\bf i}$ from the initial ground state $|g\rangle$ of the Hamiltonian $\hat{H}_{\rm d}$. The total
$e_g$ electron numbers calculated for our $16 \times 16$ cluster in the absence of the core hole are 0.87
at the nominal Mn$^{3+}$ site and 0.13 at the nominal Mn$^{4+}$ site for the parameter set with $t_0$ = 0.9~eV,
indicating a difference of 0.74 in charge density between these two sites. We note that these numbers
should not be directly compared with the local density approximation theory results or resonant
x-ray scattering results, because our effective Mn $3d$ $e_g$ states are not pure atomic Mn orbital
states but combinations of atomic Mn and O orbitals.~\cite{Herrero10} A proper comparison is described in
Appendix~\ref{sec:app-eh-number}, which shows that the electron numbers in our model are consistent 
with local density approximation and resonant x-ray scattering results. 
It is found that most of these electrons occupy the lower Jahn-Teller level $\eta = -$,
approximately $\sqrt{3}(x^2-z^2)/2$ or $\sqrt{3}(y^2-z^2)/2$ orbital at the Mn$^{3+}$ site and
the $(3z^2-r^2)/2$ orbital at the Mn$^{4+}$ site, with spin parallel to $t_{2g}$ spin at each site, consistent
with the orbital ordering proposed in Ref.~\onlinecite{Zeng08}.
\begin{figure}
    \centering
    \includegraphics[width=\hsize,clip]{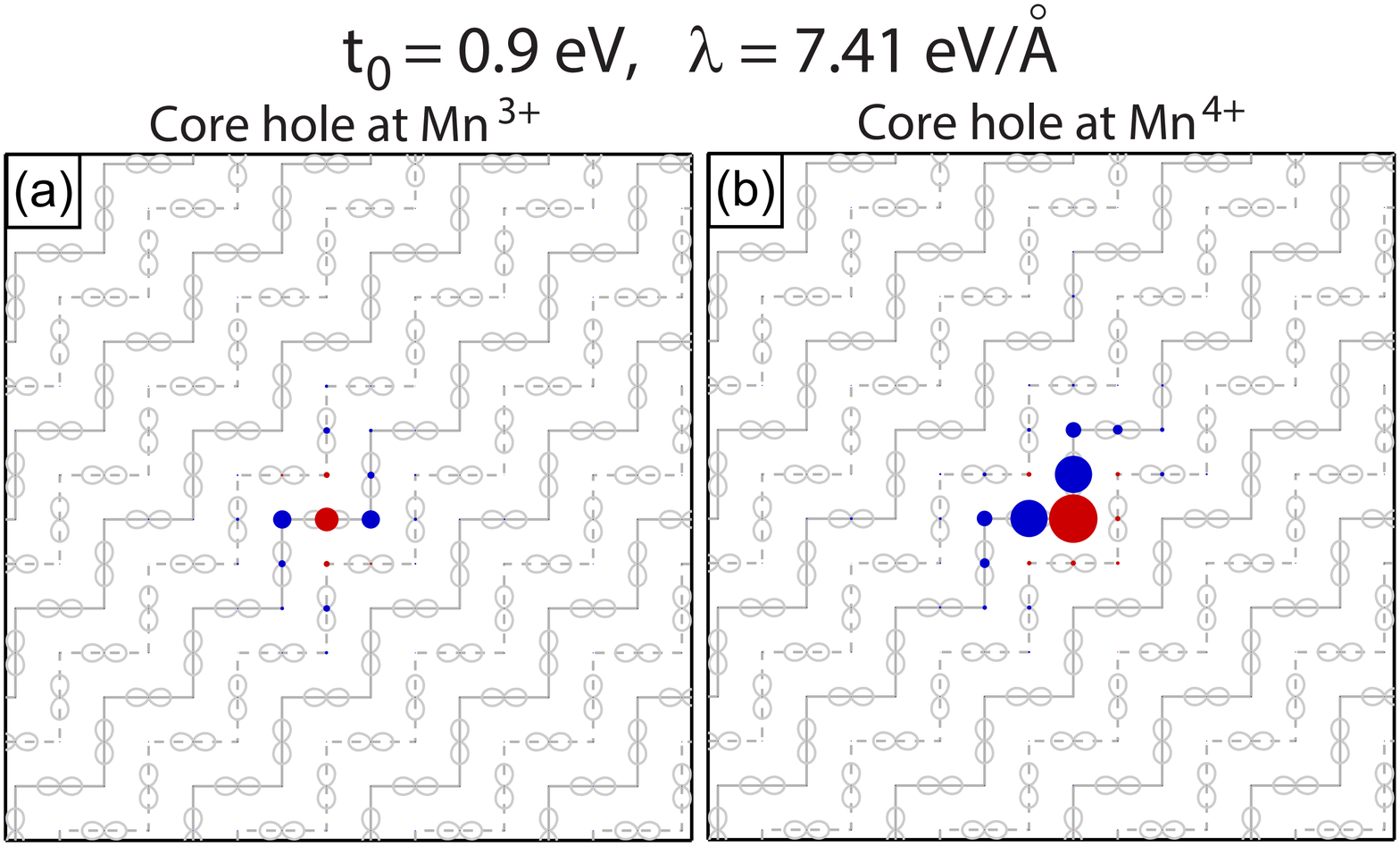}
    \caption{(Color online)
    Screening configuration in real space for $t_0=0.9$~eV with a core hole at a (a) Mn$^{3+}$, and
    (b) Mn$^{4+}$ site. The volume of the red and blue spheres is proportional to the excited electron and
    hole numbers, respectively. The excited electron number at the core-hole site (the site with the largest
    red sphere) is 0.11 for (a) and 0.92 for (b).
    } \label{fig:i-map-t0.9}
\end{figure}
\begin{figure}
    \centering
    \includegraphics[width=\hsize,clip]{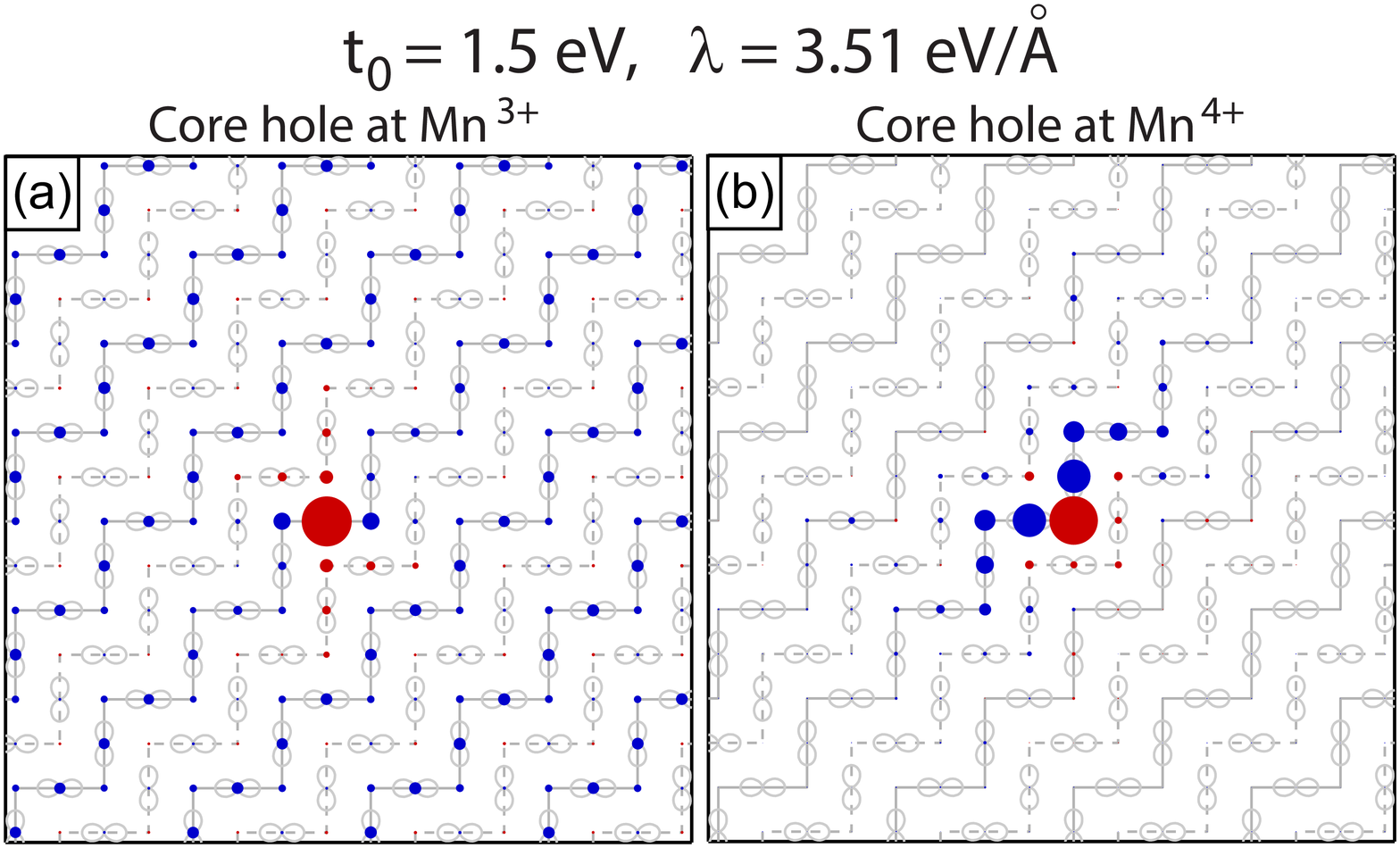}
    \caption{(Color online)
    Screening configuration in real space for $t_0=1.5$~eV with a core hole at a (a) Mn$^{3+}$, and
    (b) Mn$^{4+}$ site, similar to Fig.~\ref{fig:i-map-t0.9}. The excited electron number at the core-hole
    site is 1.02 for (a) and 0.91 for (b).
    } \label{fig:i-map-t1.5}
\end{figure}

In the intermediate state, these electron numbers change to screen the core hole. The changes in electron
number, that is, excited electron and hole numbers, are shown in Fig.~\ref{fig:i-map-t0.9} for $t_0$ = 0.9~eV
and in Fig.~\ref{fig:i-map-t1.5} for $t_0$ = 1.5~eV, where the volumes of red and blue spheres are
proportional to the excited electron and hole numbers, respectively. Note that these changes in electron
and hole numbers are consistent with the excited electron and hole distributions along the energy axis
reported in Figs.~\ref{fig:dos-nenh-t0.9}(d), \ref{fig:dos-nenh-t0.9}(e), \ref{fig:dos-nenh-t1.5}(d) and
\ref{fig:dos-nenh-t1.5}(e). The site with the largest electron number in each panel corresponds to
the core-hole site. The gray solid and dashed lines in the background represent the zigzag chain with
$t_{2g}$ spin $\uparrow$ and $\downarrow$, respectively. For the $t_0$ = 0.9~eV case,
Figs.~\ref{fig:i-map-t0.9}(a) and \ref{fig:i-map-t0.9}(b) show that the excited electrons are mostly
confined right at the core-hole site. The localization of the electrons in the intermediate state leads
to the relatively broad electron distribution, $n^e_{\rm fin \uparrow}(\varepsilon)$, along the energy axis
in the upper panels of Figs.~\ref{fig:dos-nenh-t0.9}(d) and \ref{fig:dos-nenh-t0.9}(e). Comparison of
the largest solid red spheres in Figs.~\ref{fig:i-map-t0.9}(a) and \ref{fig:i-map-t0.9}(b) shows that more
screening electrons accumulate at the core-hole site when the core hole is created at the Mn$^{4+}$ site
(0.92~electron) than at the Mn$^{3+}$ (0.11~electron). This result can be understood from the orbital
ordering pattern: Initially, the Mn$^{4+}$ site has less $e_g$ electrons on the site itself but more
electrons at its nearest neighbor Mn sites along the zigzag chain with orbitals pointing toward
the Mn$^{4+}$ site compared to the Mn$^{3+}$ site, which allows the core hole at Mn$^{4+}$ sites to
attract more electrons. Hole distribution in Figs.~\ref{fig:i-map-t0.9}(a) and \ref{fig:i-map-t0.9}(b)
show that these screening electrons are mostly from the nearest neighbors along the zigzag chain,
accounting for about 90\% of the total hole number. The results show that even though the hole excitation
is not as localized as the electron excitation and $n^h_{\rm fin \uparrow}(\varepsilon)$ is sharper
along the energy axis than  $n^e_{\rm fin \uparrow}(\varepsilon)$ in Fig.~\ref{fig:dos-nenh-t0.9}(d)
and \ref{fig:dos-nenh-t0.9}(e), the holes are still tightly bound to the core-hole site forming
an exciton-like electron-hole-pair state.

The situation changes for the case of large electron hopping $t_0$ = 1.5~eV. Figure~\ref{fig:i-map-t1.5}(a)
shows the electron-hole excitations for a core hole at a Mn$^{3+}$ site. In this case, the hole distribution
becomes delocalized, and only about 8\% of the hole is localized within the nearest neighbors of the core-hole
site, while the majority of the hole is delocalized along the zigzag chains with the same spin direction
as the core-hole site, consistent with the result in Figs.~\ref{fig:dos-nenh-t1.5}(b) and
\ref{fig:dos-nenh-t1.5}(d). The hole number does not decay with the distance from the core-hole site,
indicating qualitatively different screening dynamics. Even for the case with a core hole at a Mn$^{4+}$
site, Fig.~\ref{fig:i-map-t1.5}(b), the hole distribution spreads to further neighbors along the zig-zag
chain, reflecting the tendency toward delocalization.

As we shall see, such screening patterns in real space can be related to the variation of the RIXS
intensity in reciprocal space. This will be discussed in Sec.~\ref{ssec:rixs-intesity}.

\subsection{Calculated K-edge RIXS spectrum and comparison with experimental data}

We calculate the RIXS intensity, $I(\omega,{\bf Q})$ according to Eq.~(\ref{eq:IRIXS}). To make a comparison
between the calculated results and the experimental data, we first examine the experimental data more
closely. In addition to the momentum-dependent RIXS peak at around 2~eV, the experimental RIXS spectrum
in Fig.~\ref{fig:exp-vs-theor} shows momentum-independent spectral weight, in particular above 3~eV.
The shape of the experimental RIXS spectrum, particularly with small in-plane wavevector changes, such as
$Q_x=Q_y=-0.03\times\frac{2\pi}{a}$ in Fig.~\ref{fig:exp-vs-theor}(a), indicates that the RIXS spectral
weight above 3~eV may have the same origin as the 4--5~eV O $2p$-Mn $3d$ transition observed in optical
experiments in related manganites.\cite{Jung00} Based on this assumption, we model the experimental RIXS
spectrum with a momentum-independent peak, shown in dashed lines in Fig.~\ref{fig:exp-vs-theor}, centered
at 4.5~eV and with a half-width at half-maximum of 1.5~eV similar to the optical peak, and a momentum-dependent
Mn $3d$-$3d$ peak around 2~eV, calculated from Eq.~(\ref{eq:IRIXS}) using $\hat{H}_{\rm d}$ and
$\hat{H}_{{\rm total},{\bf i}_c}$.
Here, although it does not affect the results much,
we use the RIXS intensity averaged over twin domains in the crystal
with zigzag chains along the [$110$] and [$\bar110$] directions,
$I_{\rm avg}(\omega,(Q_x,Q_y))=[I(\omega,(Q_x,Q_y))+I(\omega,(-Q_x,Q_y))]/2$.
The results in Fig.~\ref{fig:exp-vs-theor} demonstrate a reasonable
agreement between the calculated spectra shown in solid lines and the experimental data shown in symbols,
considering the experimental noise. The 4.5~eV peak has a substantial tail even in the range of 1--3~eV.
Such momentum-independent tails have also been observed in bilayer manganites.~\cite{Weber10}
\begin{figure}
    \centering
    \includegraphics[width=\hsize,clip]{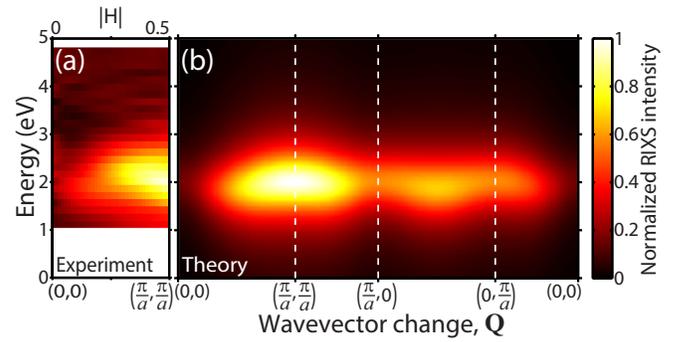}
    \caption{(Color online)
    (a) Experimental RIXS intensity. Here we take experimental data of Fig.~\ref{fig:exp-vs-theor},
    subtract the 4.5~eV peak, and fit the resulting data with a fifth order polynomial, which is
    plotted as a contour plot.
    (b) Contour plot of the calculated RIXS intensity for the $t_0=0.9$~eV case along the chosen path in
    reciprocal space.
    } \label{fig:contour-path-t0.9}
\end{figure}
\begin{figure}
    \centering
    \includegraphics[width=\hsize,clip]{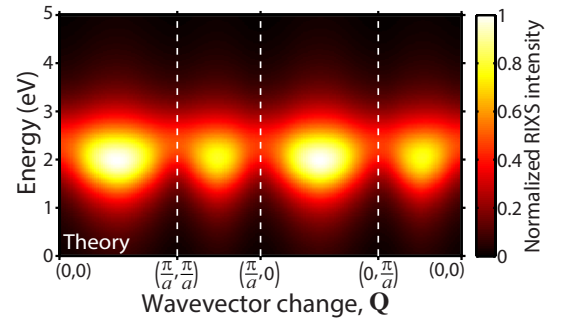}
    \caption{(Color online)
    Contour plot of the calculated RIXS intensity for $t_0=1.5$~eV along the chosen path in reciprocal space.
    } \label{fig:contour-path-t1.5}
\end{figure}

To compare just the 2~eV peak between the theory and experiment, we subtract the 4.5~eV peak from
the experimental data and plot the intensity, as a contour plot in the plane of energy and
$Q_x = Q_y = H_{\rm ex}\frac{2\pi}{a}$ in Fig.~\ref{fig:contour-path-t0.9}(a). This clearly shows
the momentum dependence of the intensity of the 2~eV peak. The calculated RIXS intensity in
Fig.~\ref{fig:contour-path-t0.9}(b) for $t_0$ = 0.9~eV shows good agreement with experimental data.
In contrast, in Fig.~\ref{fig:contour-path-t1.5}, we show the calculated RIXS spectrum for $t_0$ = 1.5~eV,
which is not consistent with the experimental data.
\begin{figure}
    \centering
    \includegraphics[width=\hsize,clip]{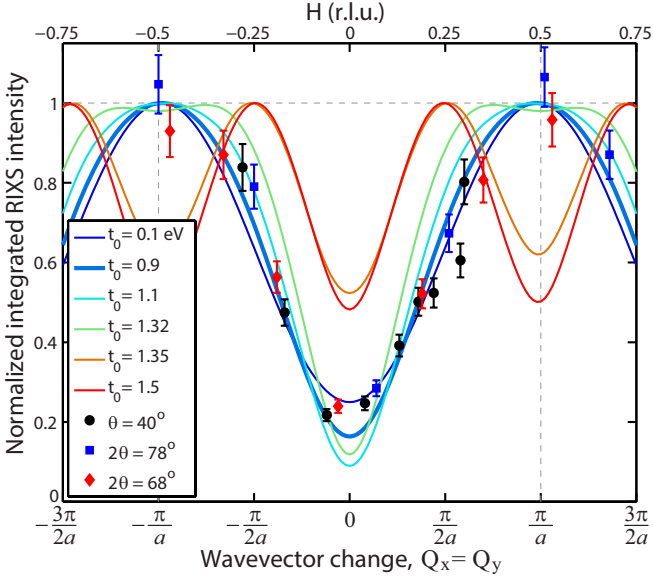}
    \caption{(Color online)
    K-edge RIXS intensity integrated from 1 to 3~eV as a function of wavevector change, normalized with
    respect to the maximum. Symbols with error bars represent experimental data. Lines represent the results
    from the theory, for several $t_0$ cases for comparison.
    } \label{fig:k-lines}
\end{figure}

\subsection{Energy-integrated RIXS intensity in reciprocal space}\label{ssec:rixs-intesity}

To make a more quantitative comparison between theory and experiment, we integrate the spectrum of
the 2~eV peak from 1~eV to 3~eV after subtracting 4.5~eV peak, for both theory and experiment.
The results are shown in Fig.~\ref{fig:k-lines} along the diagonal direction in reciprocal space,
in which the theoretical results for the various parameter sets, and the experimental data are normalized
with respect to the maximum integrated intensity. The parameter sets used for the calculations
are $(t_0,\lambda)$ = (0.1~eV, 10.79~eV/{\AA}), (0.9~eV, 7.41~eV/{\AA}), (1.1~eV, 4.81~eV/{\AA}),
(1.32~eV, 3.76~eV/{\AA}), (1.35~eV, 3.73~eV/{\AA}), and (1.50~eV, 3.51~eV/{\AA}), chosen to keep
the peak at around 2~eV. All other parameter values are unchanged. The experimental data in Fig.~\ref{fig:k-lines}
shows that the integrated RIXS intensity increases 4--5 times as the wavevector ${\bf Q}$ varies from
${\bf Q}=(0,0)$ to ${\bf Q}=(\frac{\pi}{a},\frac{\pi}{a})$. Considering fluctuations in experimental
data, the theoretical results for $t_0$ = 0.1, 0.9, 1.1~eV, all of which have exciton-like screening
electron-hole excitations similar to Fig.~\ref{fig:i-map-t0.9}, fit the experimental data reasonably
well. In contrast, the theoretical results for $t_0$ = 1.35 and 1.5~eV, all of which have delocalized
hole excitations similar to Fig.~\ref{fig:i-map-t1.5}, are qualitatively different from experimental
data with maximum intensity at around $(\pm\frac{\pi}{2a},\pm\frac{\pi}{2a})$ instead of
$(\pm\frac{\pi}{a},\pm\frac{\pi}{a})$. This provides an upper limit of about 1.2~eV for the value of $t_0$.

This analysis indicates that, irrespective of the details of the model Hamiltonian and the particular
parameter values, the rapid increase of the RIXS intensity with a maximum at $(\pm\frac{\pi}{a},\pm\frac{\pi}{a})$
observed in the experiment is indicative of highly localized screening dynamics~\cite{note-PiPi} in
La$_{0.5}$Sr$_{1.5}$MnO$_4$, i.e., screening that is more like Fig.~\ref{fig:i-map-t0.9} than that of
Fig.~\ref{fig:i-map-t1.5}.

\subsection{Periodicity of K-edge RIXS spectrum in reciprocal space for charge-orbital-spin ordered manganites}\label{ssec:rixs-periodic}

As mentioned in Sec.~\ref{ssec:rixs-kspace}, in earlier studies of La$_2$CuO$_4$, it was shown that
the observed K-edge RIXS spectrum reflects the periodicity of the lattice and is a function of the reduced
wavevector ${\bf q}$ within the first Brillouin zone, defined as ${\bf Q}={\bf q}+{\bf K}$, where ${\bf Q}$
is the total x-ray wavevector change and ${\bf K}$ is a reciprocal lattice vector.~\cite{Kim07}
\begin{figure}
    \centering
    \includegraphics[width=\hsize,clip]{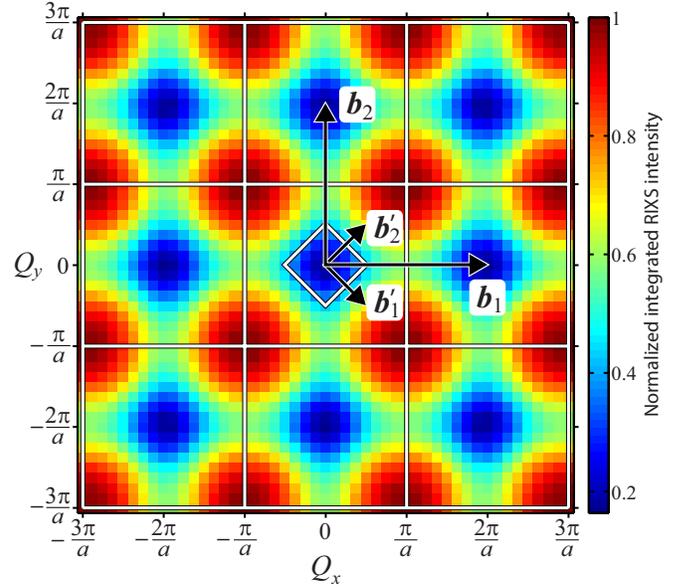}
    \caption{(Color online)
    K-edge RIXS intensity integrated from 1~eV to 3~eV, shown in a large reciprocal space for the
    $t_0=0.9$~eV case. ${\bf b}'_1$ and ${\bf b}'_2$ represent the primitive reciprocal lattice vectors
    of charge-orbital-spin ordered MnO$_2$ plane of Fig.~\ref{fig:ce-order}. ${\bf b}_1$ and ${\bf b}_2$
    are the primitive reciprocal lattice vectors for MnO$_2$ lattice without ordering.
    } \label{fig:k-map-ext-t0.9}
\end{figure}
\begin{figure}
    \centering
    \includegraphics[width=\hsize,clip]{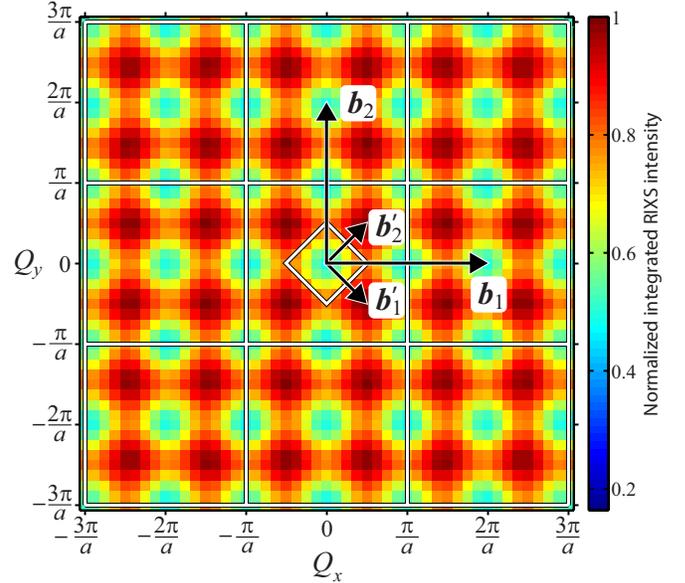}
    \caption{(Color online)
    Similar figure as Fig.~\ref{fig:k-map-ext-t0.9} for the $t_0=1.5$~eV case.
    } \label{fig:k-map-ext-t1.5}
\end{figure}

The experimental data presented in this paper clearly indicates that such periodicity is not present for
La$_{0.5}$Sr$_{1.5}$MnO$_4$. The measured RIXS intensity, as well as the calculated RIXS intensity, seen in
Figs.~\ref{fig:contour-path-t0.9} and \ref{fig:k-lines}, increases continuously past the boundary of the first
Brillouin zone at $(\frac{\pi}{4a},\frac{\pi}{4a})$. As discussed in Sec.~\ref{ssec:rixs-kspace},
the periodicity seen in Ref.~\onlinecite{Kim07} applies only to solids with one core-hole site per unit cell,
such as La$_2$CuO$_4$. 
For solids with multiple core hole sites per unit cell due to the ordering of spin, charge, orbital or local lattice distortions,
the periodicity in K-edge RIXS spectrum follows the periodicity of the lattice without ordering,
like the square Mn-site lattice in La$_{0.5}$Sr$_{1.5}$MnO$_4$,
not the periodicity of the actual lattice with ordering. 
Our numerical calculations in
Figs.~\ref{fig:k-map-ext-t0.9} and \ref{fig:k-map-ext-t1.5} confirm such periodicity in
La$_{0.5}$Sr$_{1.5}$MnO$_4$: Fig.~\ref{fig:k-map-ext-t0.9} shows energy-integrated RIXS intensity calculated
for the $t_0$ = 0.9~eV case in a larger region of reciprocal space of $-\frac{3\pi}{a}<Q_x\leq\frac{3\pi}{a}$
and $-\frac{3\pi}{a}<Q_y\leq\frac{3\pi}{a}$. The diamond around $(0,0)$ is the actual first Brillouin zone of
the spin, charge and orbital ordered structure, whereas the outer square domain
$-\frac{\pi}{a}<Q_x\leq\frac{\pi}{a}$ and $-\frac{\pi}{a}<Q_y\leq\frac{\pi}{a}$, denotes the first Brillouin
zone of the lattice without ordering. 
It is evident that RIXS spectrum does not exhibit periodicity with
respect to the actual primitive reciprocal lattice vectors, ${\bf b}'_1$ and ${\bf b}'_2$, shown in
Fig.~\ref{fig:k-map-ext-t0.9}, but rather shows periodicity with respect to primitive reciprocal lattice
vectors of the lattice without ordering, ${\bf b}_1$ and ${\bf b}_2$, shown in Fig.~\ref{fig:k-map-ext-t0.9}.
Even though it is only a single data point, the experimental data at around $Q_x=Q_y=0.68\times\frac{2\pi}{a}$
in Fig.~\ref{fig:k-lines} is consistent with such periodicity. Further experimental data for
$|Q_x|>\frac{\pi}{a}$, $|Q_y|>\frac{\pi}{a}$ are required for the verification of this periodicity. We note
that careful examination of Fig.~\ref{fig:k-map-ext-t0.9} reveals that the periodicity is only approximate.
The small displacement of the Mn$^{4+}$ ions of 0.0265~{\AA} along the diagonal direction from the ideal
square lattice,~\cite{Zeng08} included in our calculations, makes the Mn sites deviate slightly
from the exact square lattice. Figure~\ref{fig:k-map-ext-t1.5} shows that, even for the $t_0$ = 1.5~eV case,
the periodicity still does not follow the actual reciprocal lattice, even though the RIXS intensity oscillates
more rapidly in reciprocal space.
\begin{figure}
    \centering
    \includegraphics[width=\hsize,clip]{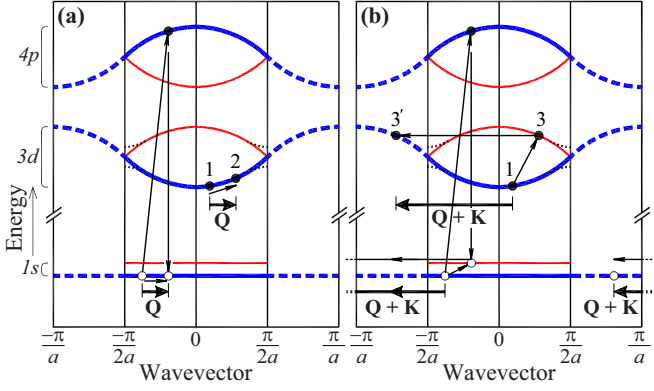}
    \caption{(Color online)
    Schematic diagrams of a band structure of a one-dimensional system with a lattice constant $a$, before
    [thick (blue) lines, both solid and dashed] and after [solid lines, both thick (blue) and thin (red)]
    {\it artificial} unit cell doubling. The thin black dotted lines in the $3d$ bands represent the modification of the band structure
    after {\it real} unit cell doubling due to charge-orbital ordering. Panels (a) and (b) represent the RIXS
    processes that result in x-ray wavevector transfers of ${\bf Q}$ and ${\bf Q + K}$, respectively, with
    ${\bf K} = -\frac{\pi}{a}$, a reciprocal lattice vector for the lattice with the doubled unit cell.
    } \label{fig:periodic}
\end{figure}

We next provide a physical explanation of why the K-edge RIXS spectrum from charge-orbital ordered manganites
follows the periodicity of the Brillouin zone in the absence of charge-orbital order. We first consider
\emph{artificial} doubling of the unit cell for a one-dimensional chain with interatomic distance $a$.
In Fig.~\ref{fig:periodic}, thick (blue) lines, both solid and dashed, and solid lines, both thick (blue) and thin (red),
show a schematic
diagram of the band structure before and after the artificial unit cell doubling, respectively.
The arrows between $1s$ and $4p$ bands represent the core hole creation and annihilation by x-rays.
Due to the interaction with
the core-hole,
electron-hole pairs can be excited into the valence shell, as indicated by the arrows within the $3d$ band.
Excitations with a transferred crystal wavevector ${\bf Q}$ from
the state $1$  can be either an intraband transition to the state $2$ shown in Fig.~\ref{fig:periodic}(a)
or an interband transition to the state 3 shown in Fig.~\ref{fig:periodic}(b).
However, the reduced Brillouin zone
$[-\frac{\pi}{2a}, \frac{\pi}{2a}]$ is an artificial construction and should give equivalent results to
the real Brillouin zone $[-\frac{\pi}{a},\frac{\pi}{a}]$, in which the state $3'$ in Fig.~\ref{fig:periodic}(b)
should be considered instead of the state $3$.
Therefore, the intraband and interband transitions correspond
to wavevector transfers of ${\bf Q}$ and ${\bf Q+K}$, respectively, where ${\bf K}$ is $-\frac{\pi}{a}$,
a reciprocal lattice vector of the doubled unit cell. This implies that they occur at two distinct wavevector transfers
and should be distinguishable. 
The underlying reason is that
not only the valence bands are backfolded due to the doubling of unit cell, but also the core level bands.
An interband (intraband) transition in the valence band also leads to an interband (intraband) transition
in the core level band, as shown in arrows in the $1s$ band in Fig.~\ref{fig:periodic}. 
We now consider {\it real} unit cell doubling due to charge-orbital order. For
a finite but small charge-orbital order, the band structure would be modified mostly near the Brillouin zone
boundary, as represented by the thin black dotted lines in the $3d$ bands 
in Fig.~\ref{fig:periodic}, and the RIXS that
involves states far from the zone boundary, such as the states $1$, $2$ and $3$, should not dramatically change
from those in the absence of charge-orbital order. Therefore, the K-edge RIXS spectrum $I(\omega,{\bf Q})$ and
$I(\omega,{\bf Q+K})$ would remain different even after charge-orbital ordering. 
Obviously, the situation
becomes more complex when the charge-orbital order becomes stronger leading to a further mixing of the bands.
However, the schematic figure explains why the periodicity for charge-orbital ordered manganites occurs with
the Brillouin zone in the absence of such order.

\section{Discussion}\label{sec:discussion}

\begin{figure}
    \centering
    \includegraphics[width=\hsize,clip]{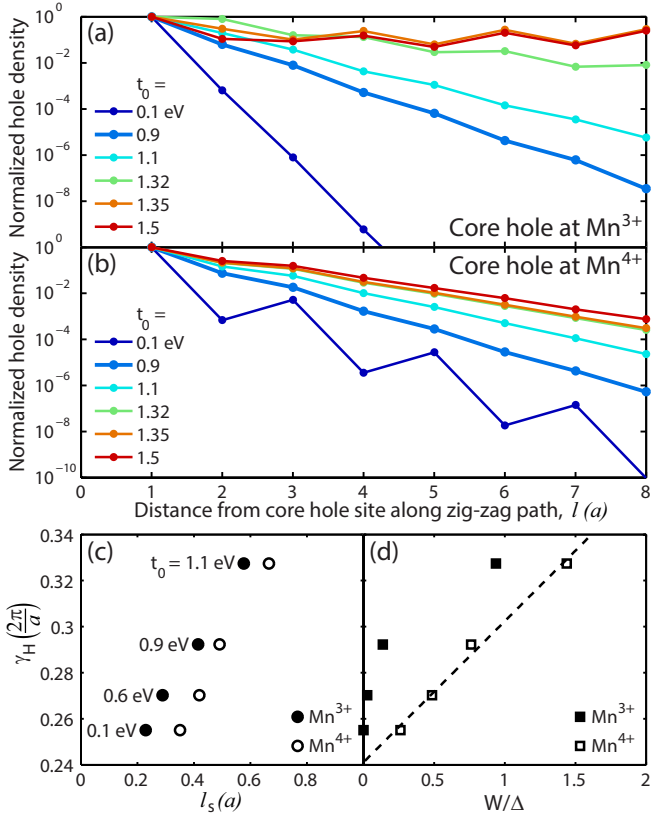}
    \caption{(Color online)
    (a) and (b): Semi-log plot of the excited hole number normalized 
    to its maximum versus the distance 
    from the core-hole site for a Mn$^{3+}$, and a Mn$^{4+}$ core hole site, respectively.
    (c) The width of the RIXS peak at $(\pm\frac{\pi}{a},\pm\frac{\pi}{a})$ in reciprocal space,
    $\gamma_H$, found from Fig.~\ref{fig:k-lines}, versus screening cloud size in real space $l_s$,
    found from (a) and (b), for various $t_0$ values.
    (d) $\gamma_{\rm H}$ versus the ratio between the occupied band width $(W)$ and the in-gap bound
    state energy $(\Delta)$ from the occupied band edge, $W/\Delta$.
    } \label{fig:holedensity-multi}
\end{figure}

In this section, we discuss what insights the RIXS spectrum in reciprocal space can provide on the screening
dynamics. Figures~\ref{fig:holedensity-multi}(a) and \ref{fig:holedensity-multi}(b) show $n_h/n_h^{\max}$,
that is the hole density normalised to its maximum, versus distance $l$ from the core-hole site along the
zig-zag chain for the case of core hole at a Mn$^{3+}$ and Mn$^{4+}$ site, respectively, in a semi-logarithmic
plot. For $t_0\leq1.3$~eV of Fig.~\ref{fig:holedensity-multi}(a) and all $t_0$ of
Fig.~\ref{fig:holedensity-multi}(b), the hole density decreases exponentially and thus can be fit to
$n_h/n_h^{\max}\propto\exp(-l/l_s)$, where $l_s$ can be interpreted as the size of the screening cloud.
We find that the sizes of the screening clouds are approximately $0.4$ and $0.5$ interatomic distances for
the Mn$^{3+}$ and Mn$^{4+}$ sites, respectively, for $t_0$ = 0.9~eV, and become larger, as $t_0$ increases,
as shown along the horizontal axis in Fig.~\ref{fig:holedensity-multi}(c), which includes the result for
$t_0 = 0.6$~eV additionally.

We next look for a correlation between the size of the screening cloud in real space and the features of
the K-edge RIXS spectrum in reciprocal space. We define $\gamma_H$ as the half-width-at-half-maximum
in reciprocal space of the energy-integrated RIXS peak for $t_0\leq1.3$~eV in Fig.~\ref{fig:k-lines}. The plot
of $\gamma_H$ versus $l_s$ is displayed in Fig.~\ref{fig:holedensity-multi}(c), which shows a linear
correlation. Therefore, the width of the energy-integrated RIXS peak in reciprocal space in
Fig.~\ref{fig:k-lines} can be used to estimate the size of the screening hole cloud.
This is an important result with general implications.

The size of
the screening hole cloud depends on the competition between hole hopping and hole binding energies,
which can be parameterized in terms of $W$, the width of the occupied band, and $\Delta$, the energy
difference between the top of the occupied band and the unoccupied bound state within the gap, or,
the hole binding energy. Therefore, from the connection between $l_s$ and $\gamma_H$ established above,
we examine whether the width of the RIXS peak $\gamma_H$ can also provide insight on the ratio $W/\Delta$.
The $\gamma_H$ versus $W/\Delta$ plot in Fig.~\ref{fig:holedensity-multi}(d) confirms a positive
correlation between these quantities, in particular a linear correlation for the case of core hole at
Mn$^{4+}$ site. The results indicate that the width of the RIXS peak can be a measure of the size of
the exciton-like screening cloud in real space, and a measure of the ratio between the occupied
band width and the hole binding energy.

Recently, K-edge RIXS spectra for bilayer manganites La$_{2-2x}$Sr$_{1+2x}$Mn$_{2}$O$_{7}$ with
$x$ = 0.36 and 0.5 have been reported by Weber~{\it et al.}~\cite{Weber10} Although not as pronounced
as our results for single layered manganites, Ref.~\onlinecite{Weber10} shows an increase of the 2~eV
peak intensity in A-type and CE-type antiferromagnetic LaSr$_{2}$Mn$_{2}$O$_{7}$, as the x-ray wavevector
transfer increases from $(0,0,Q_z)$ to $(\frac{\pi}{a},\frac{\pi}{a},Q_z)$.
In the context of our work above, such results can be interpreted as
the formation of an exciton-like screening cloud in bi-layer manganites, the size of which is
likely larger than that for the single-layer manganites discussed above, considering the less-pronounced increase of
the 2~eV peak intensity.

Finally, we note that Semba~\textit{et al.}~calculated the K-edge RIXS spectrum for LaMnO$_3$, based on
the Keldysh-type Green's function formalism.\cite{Semba08} The results in Ref.~\onlinecite{Semba08}
show about a 10\% increase of the 2~eV peak from $(0,0,0)$ to $(1,0,0)$ in their choice of $x$ and $y$ axes,
which are equivalent to ${\bf Q}=(0,0,0)$ to ${\bf Q}=(\frac{\pi}{a},\frac{\pi}{a},0)$ in our notation.
The results again can be interpreted as the formation of exciton-like screening clouds, consistent
with our calculations, even though the core-hole state for the intermediate eigenstates is chosen
as delocalised in Ref.~\onlinecite{Semba08}. The periodicity of the RIXS spectrum discussed in
Sec.~\ref{ssec:rixs-kspace} and Sec.~\ref{ssec:rixs-periodic} was also identified in
Ref.~\onlinecite{Semba08}, even though the association of such periodicity with the approximate
square lattice of the core-hole sites was not specifically mentioned.

\section{Summary}\label{sec:summary}

We have presented a formalism to calculate the K-edge RIXS spectra in transition metal oxides based on
tight-binding Hamiltonians and a local $1s$--$3d$ Coulomb interaction, in which the choice of intermediate
eigenstates with a completely localized $1s$ core hole allows for the interpretation of the data in terms
of screening dynamics in real space. 
We have also found that the periodicity of K-edge RIXS spectrum
follows the reciprocal lattice vectors of the lattice without the ordering
of spin, charge, orbital, or local lattice distortions, 
rather than the reciprocal lattice vectors of the actual lattice with such ordering.

We have applied our formalism to the highly-momentum-dependent K-edge RIXS spectrum observed for
La$_{0.5}$Sr$_{1.5}$MnO$_4$ in CE-type spin-orbital-structure ordering. It is found that the sharp increase
of the 2~eV peak intensity from the center toward the corner of the first Brillouin zone of
the lattice without ordering is an indication of a highly localized screening cloud in
La$_{0.5}$Sr$_{1.5}$MnO$_4$ with a typical size of 0.4--0.5 Mn-Mn distances.
We also showed that there exists a positive correlation between the width of the energy-integrated K-edge RIXS
intensity peak in reciprocal space, the size of the exciton-like screening cloud in real space, and
the ratio between occupied band width and hole binding energy.

The analysis in this paper was performed for the case of an intermediate strength core-hole potential, that
is $U_{\rm core}$ comparable to the $3d$ electron bandwidth. In fact, this is appropriate for most transition
metal oxides and therefore the present approach should have general applicability. One of the important results
is that this approach highlights the connection between K-edge RIXS and the impurity problem in strongly
correlated electron systems, and we show that this technique is a new probe of momentum-dependent screening
dynamics of localized impurities.

\section*{Acknowledgements}

We thank D. Casa, D. Prabhakaran, A.~T.~Boothroyd and H.~Ding for their invaluable support in experiments.
The collaborations between T.F.S., K.H.A., and M.v.V. were supported by the Computational Materials and
Chemical Science Network under Grants No.DE-FG02-08ER46540 and No.DE-SC0007091. K.H.A. was further
supported by 2012 and 2013 Argonne X-ray Science Division Visitor Program. M.v.V. was supported by the US Department
of Energy, Office of Basic Energy Sciences, Division of Materials Sciences and Engineering under Award
No.DE-FG02-03ER46097 and NIU Institute for Nanoscience, Engineering, and Technology. The work at Brookhaven
was supported by the US Department of Energy, Division of Materials Science, under Contract
No. DE-AC02-98CH10886. Work at Argonne National Laboratory and use of the Advanced Photon Source was
supported by the US DOE, Office of Science, Office of Basic Energy Sciences, under Contract
No. DE-AC02-06CH11357.

\appendix

\section{RIXS formula derivation}\label{sec:app-rixs}

As explained in Sec.~\ref{ssec:rixs-deriv}, we get the following formula from the Kramers-Heisenberg
formula, Eq.~(\ref{eq:KH}), in the limit of completely localized core hole,
\begin{eqnarray}
    &&I(\omega,{\bf k},{\bf k}',\pmb{\epsilon},\pmb{\epsilon}') \propto  \nonumber \\
    &&\sum_{f} \left|
        \sum_{\bf R} \sum_{\bf d} \sum_{n^{{\bf R}+{\bf d}}}
        \frac{ \langle f| \mathcal{D'}^{\dag}|n^{{\bf R}+{\bf d}} \rangle
        \langle n^{{\bf R}+{\bf d}}| \mathcal{D}|g \rangle }
        {E_g+\hbar \omega_{\bf k}-E_{n^{\bf d}} + i \Gamma_{n^{\bf d}}}
    \right|^2 \times \nonumber \\
    &&\delta(E_g - E_f + \hbar\omega),
\end{eqnarray}
where $|n^{{\bf R}+{\bf d}} \rangle$ represents the intermediate energy eigenstate with the core hole
at a site ${\bf R}+{\bf d}$ within the unit cell at a lattice point ${\bf R}$. Further applying the dipole
approximation to the K-edge scattering amplitude, we obtain
\begin{eqnarray}
    &&\langle f| \mathcal{D'}^{\dag}|n^{{\bf R}+{\bf d}} \rangle
        \langle n^{{\bf R}+{\bf d}}| \mathcal{D}|g \rangle
        =  e^{-i ({\bf k}'-{\bf k})\cdot ({\bf R}+{\bf d})} \times \nonumber \\
    &&  \pmb{\epsilon}' \cdot
        \langle f | {\bf r} | n^{{\bf R}+{\bf d}} \rangle \ \
        \pmb{\epsilon} \cdot
        \langle n^{{\bf R}+{\bf d}} | {\bf r} | g \rangle.
    \label{eq:D-expans1}
\end{eqnarray}

Two many-body states $|\Psi^{0} \rangle$ and $|\Psi^{{\bf R}} \rangle$, which have total momentum
$\hbar \pmb{\kappa}$ and identical wave functions in two different coordinate systems with
the coordinates for $|\Psi^{{\bf R}} \rangle$ shifted with respect to the coordinates for $|\Psi^{0} \rangle$
by ${\bf R}$, are related to each other by a phase factor $|\Psi^{{\bf R}} \rangle$ =
$e^{-i\pmb{\kappa} \cdot {\bf R}} |\Psi^{0} \rangle$. Assuming that $| g \rangle$ and $| f \rangle$
have net momenta of zero and $\hbar {\bf k}_f$, respectively, we obtain the following relation,~\cite{note-R}
\begin{eqnarray}
    \langle f | {\bf r} | n^{{\bf R}+{\bf d}} \rangle &=&
        e^{-i {\bf k}_f \cdot {\bf R}}
        \langle f | {\bf r} | n^{\bf d} \rangle, \label{eq:D-expans2} \\
    \langle n^{{\bf R}+{\bf d}} | {\bf r} | g \rangle
        &=& \langle n^{\bf d} | {\bf r} | g \rangle.
    \label{eq:D-expans3}
\end{eqnarray}
Therefore, the sum over lattice vectors ${\bf R}$ for the combined factor of
$e^{-i ({\bf k}'-{\bf k}+{\bf k}_f)\cdot {\bf R}}$ from Eqs.~(\ref{eq:D-expans1})--(\ref{eq:D-expans3})
leads to the conservation of the crystal momentum $\delta ({\bf k}'-{\bf k}+{\bf k}_f+{\bf K})$, where
${\bf K}$ represents the reciprocal lattice vectors. This results in the following expression for
the RIXS intensity,
\begin{widetext}
\begin{equation}
    I(\omega,{\bf k},{\bf k}',\pmb{\epsilon},\pmb{\epsilon}') \propto
        \sum_{\bf K} \sum_{f} \left|
            \sum_{\bf d} \sum_{n^{\bf d}}
            \frac{ e^{-i ({\bf k}'-{\bf k}) \cdot {\bf d}} \
                \pmb{\epsilon}'\cdot \langle f| {\bf r} |n^{\bf d} \rangle \;
                \pmb{\epsilon}\cdot\langle n^{\bf d}| {\bf r} |g \rangle }
                {E_g+\hbar \omega_{\bf k}-E_{n^{\bf d}} + i \Gamma_{n^{\bf d}}}
        \right|^2
    \delta(E_f+\hbar\omega_{{\bf k}'}-E_g-\hbar\omega_{\bf k}) \
        \delta({\bf k}_f+{{\bf k}'}-{\bf k}+{\bf K}).
\end{equation}
\end{widetext}
By further assuming special experimental setups in which the polarization vectors $\pmb{\epsilon}'$
and $\pmb{\epsilon}$ give rise to a constant factor, as mentioned in Sec.~\ref{ssec:rixs-deriv},
and neglecting a constant factor from the dipole moment between $4p$ and $1s$ wave functions, we obtain
Eq.~(\ref{eq:KH-IRIXS}) in Sec.~\ref{ssec:rixs-deriv}.

\section{Expression of the matrix elements in the K-edge RIXS formula in terms of eigenstates
in the presence and absence of core hole}\label{sec:app-H-itok}

In general, we transform $\hat{H}_{\rm d}$ and $\hat{H}_{{\rm total},{\bf i}_c}$ into the reciprocal
space as follows,
\begin{eqnarray}
    \hat{H}_{\rm d} =
    \sum_{{\bf k},{\bf k}'} \sum_{{\bf K},{\bf K}'} \sum_{\xi,\xi'} \sum_{\sigma}
        && H^{\rm d}_{{\bf k}+{\bf K},\xi,{\bf k}'+{\bf K}',\xi',\sigma} \ \times \nonumber\\
        && d_{{\bf k}+{\bf K},\xi,\sigma}^{\dag} d_{{\bf k}'+{\bf K}',\xi',\sigma},
\end{eqnarray}
and
\begin{eqnarray}
    \hat{H}_{{\rm total},{\bf i}_c} =
    \sum_{{\bf k},{\bf k}'} \sum_{{\bf K},{\bf K}'} \sum_{\xi,\xi'} \sum_{\sigma}
        && H^{{\rm total},{\bf i}_c}_{{\bf k}+{\bf K},\xi,{\bf k}'+{\bf K}',\xi',\sigma} \ \times \nonumber\\
        && d_{{\bf k}+{\bf K},\xi,\sigma}^{\dag} d_{{\bf k}'+{\bf K}',\xi',\sigma},
\end{eqnarray}
where ${\bf k}$ and ${\bf k}'$ represent vectors within the first Brillouin zone $\Omega_{\rm 1BZ}$,
${\bf K}$ and ${\bf K}'$ the reciprocal lattice vectors within the extended ``first Brillouin zone"
$\Omega_{\rm ExZ}$ defined by the core hole site ${\bf i}$. Spin states are represented by $\sigma$, and
orbital states by $\xi$ and $\xi'$.

From the eigenstates $|l{\bf k}\sigma\rangle$ of $\hat{H}_{\rm d}$ with the wavevector
${\bf k}\in \Omega_{\rm 1BZ}$ within the $l$-th lowest energy band, and the $m$-th lowest
energy eigenstate $|m\sigma\rangle$ of $\hat{H}_{{\rm total},{\bf i}_c}$, we define
$\beta_{l{\bf k}m\sigma}$ = $\langle l{\bf k}\sigma | m\sigma\rangle$.

In the RIXS formula Eq.~(\ref{eq:IRIXS}), $\langle n^{\bf d}_{\rm low}| \underline{s}_{\bf d}^{\dagger} |g \rangle$ and
$\langle l_e {\bf k}_e l_h {\bf k}_h \sigma | \underline{s}_{\bf d} |n^{\bf d}_{\rm low} \rangle$ are found from
\begin{widetext}
\begin{equation}
    \langle n^{\bf d}_{\rm low}| \underline{s}_{\bf d}^{\dagger} |g \rangle =
    \prod_{\sigma=\uparrow\downarrow}
    \left|
    \begin{array}{cccc}
    \beta_{1{\bf k}_1 1\sigma}         & \beta_{1{\bf k}_1 2\sigma}         & \cdots & \beta_{1{\bf k}_1 \frac{N_e}{2}\sigma} \\
    \beta_{1{\bf k}_2 1\sigma}         & \beta_{1{\bf k}_2 2\sigma}         & \cdots & \beta_{1{\bf k}_2 \frac{N_e}{2}\sigma} \\
    \vdots                             & \vdots                             & \ddots & \vdots \\
    \beta_{l_h^{\rm max} {\bf k}_{N_k} 1\sigma} & \beta_{l_h^{\rm max} {\bf k}_{N_k} 2\sigma} & \cdots & \beta_{l_h^{\rm max} {\bf k}_{N_k} \frac{N_e}{2}\sigma}
    \end{array}
    \right|,
\end{equation}
\begin{equation} \label{eq:det-feh}
    \langle l_e {\bf k}_e l_h {\bf k}_h \sigma | \underline{s}_{\bf d} |n^{\bf d}_{\rm low} \rangle = 
    \left|
    \begin{array}{cccc}
    \beta_{1{\bf k}_1 1\sigma}         & \beta_{1{\bf k}_1 2\sigma}         & \cdots & \beta_{1{\bf k}_1 \frac{N_e}{2}\sigma} \\
    \beta_{1{\bf k}_2 1\sigma}         & \beta_{1{\bf k}_2 2\sigma}         & \cdots & \beta_{\sigma 1{\bf k}_2\frac{N_e}{2}} \\
    \vdots                             & \vdots                             & \ddots & \vdots \\
    \beta_{l_h''{\bf k}_h'' 1\sigma}   & \beta_{l_h''{\bf k}_h'' 2\sigma}   & \cdots & \beta_{l_h''{\bf k}_h'' \frac{N_e}{2}\sigma} \\
    \beta_{l_e{\bf k}_e 1\sigma}       & \beta_{l_e{\bf k}_e 2\sigma}       & \cdots & \beta_{l_e{\bf k}_e \frac{N_e}{2}\sigma} \\
    \beta_{l_h'''{\bf k}_h''' 1\sigma} & \beta_{l_h'''{\bf k}_h''' 2\sigma} & \cdots & \beta_{l_h'''{\bf k}_h''' \frac{N_e}{2}\sigma} \\
    \vdots                             & \vdots                             & \ddots & \vdots \\
    \beta_{l_h^{\rm max} {\bf k}_{N_k} 1\sigma}     & \beta_{l_h^{\rm max} {\bf k}_{N_k} 2\sigma}     & \cdots & \beta_{l_h^{\rm max} {\bf k}_{N_k} \frac{N_e}{2}\sigma}
    \end{array}
    \right| \times 
    \left|
    \begin{array}{cccc}
    \beta_{1{\bf k}_1 1\bar\sigma}     & \beta_{1{\bf k}_1 2\bar\sigma}     & \cdots & \beta_{1{\bf k}_1 \frac{N_e}{2}\bar\sigma} \\
    \beta_{1{\bf k}_2 1\bar\sigma}     & \beta_{1{\bf k}_2 2\bar\sigma}     & \cdots & \beta_{1{\bf k}_2 \frac{N_e}{2}\bar\sigma} \\
    \vdots                             & \vdots                             & \ddots & \vdots   \\
    \beta_{l_h^{\rm max} {\bf k}_{N_k} 1\bar\sigma} & \beta_{l_h^{\rm max} {\bf k}_{N_k} 2\bar\sigma} & \cdots & \beta_{l_h^{\rm max} {\bf k}_{N_k} \frac{N_e}{2}\bar\sigma}
    \end{array}
    \right|
\end{equation}
\end{widetext}
where $N_e$ represents the total electron number, $N_{k}$ the number of ${\bf k}$-points in $\Omega_{\rm 1BZ}$,
$l_h^{\rm max}$ the index for the highest occupied band, and $\bar\sigma = -\sigma$. In Eq.~(\ref{eq:det-feh}),
the set of band and momentum indices, ($l_h''$, ${\bf k}_h''$) and ($l_h'''$, ${\bf k}_h'''$), represent
the occupied states right before and right after the hole state represented by ($l_h$, ${\bf k}_h$) when
the eigenstates of $\hat{H}_{\rm d}$ are ordered according to the band index and momentum index.~\cite{note-beta}

\section{Expressions of the Hamiltonians in reciprocal space with and without a $1s$ core hole
for L\lowercase{a}$_{0.5}$S\lowercase{r}$_{1.5}$M\lowercase{n}O$_4$}\label{sec:app-hamilt}

In the absence of the core hole, the Hamiltonian for a single layer of La$_{0.5}$Sr$_{1.5}$MnO$_4$ has
the following form in reciprocal space,
\begin{equation}
    \hat{H}_{\rm d} = \sum_{\sigma,{\bf k} \in \Omega_{\rm 1BZ}}
        d^{\dagger}_{{\bf k}\sigma} \left(
            H^{\rm d,nonint}_{{\bf k}\sigma} + H^{\rm dd,HF}_{{\bf k}\sigma}
        \right) d_{{\bf k}\sigma},
\end{equation}
where
\begin{widetext}
\begin{eqnarray}
    d^{\dagger}_{{\bf k}\sigma} &=& (
    d^{\dagger}_{{\bf k}+{\bf K}_1,1,\sigma},
    d^{\dagger}_{{\bf k}+{\bf K}_1,2,\sigma},
    d^{\dagger}_{{\bf k}+{\bf K}_2,1,\sigma},
    d^{\dagger}_{{\bf k}+{\bf K}_2,2,\sigma},
    d^{\dagger}_{{\bf k}+{\bf K}_3,1,\sigma},
    d^{\dagger}_{{\bf k}+{\bf K}_3,2,\sigma},
    d^{\dagger}_{{\bf k}+{\bf K}_4,1,\sigma},
    d^{\dagger}_{{\bf k}+{\bf K}_4,2,\sigma},
    \nonumber \\ & &
    d^{\dagger}_{{\bf k}+{\bf K}_5,1,\sigma},
    d^{\dagger}_{{\bf k}+{\bf K}_5,2,\sigma},
    d^{\dagger}_{{\bf k}+{\bf K}_6,1,\sigma},
    d^{\dagger}_{{\bf k}+{\bf K}_6,2,\sigma},
    d^{\dagger}_{{\bf k}+{\bf K}_7,1,\sigma},
    d^{\dagger}_{{\bf k}+{\bf K}_7,2,\sigma},
    d^{\dagger}_{{\bf k}+{\bf K}_8,1,\sigma},
    d^{\dagger}_{{\bf k}+{\bf K}_8,2,\sigma})
\end{eqnarray}
with ${\bf K}_1$, ${\bf K}_2$, ${\bf K}_3$, ${\bf K}_4$, ${\bf K}_5$, ${\bf K}_6$, ${\bf K}_7$, and ${\bf K}_8$
representing $(0,0)$, $(\frac{\pi}{a},0)$, $(0,\frac{\pi}{a})$, $(\frac{\pi}{a},\frac{\pi}{a})$,
$(-\frac{\pi}{2a},-\frac{\pi}{2a})$, $(\frac{\pi}{2a},-\frac{\pi}{2a})$ , $(-\frac{\pi}{2a},\frac{\pi}{2a})$,
and $(\frac{\pi}{2a},\frac{\pi}{2a})$,
respectively,
\begin{equation}
    H^{\rm d,nonint}_{{\bf k}\sigma} =
    \left(
    \begin{array}{cccccccc}
    M_1+W_{3u} & -G_{\sigma} & G_{\sigma} & W_{1s}+W_{3s} & W_{2s} & G_{\sigma} & G_{\sigma} & W_{2s} \\
    -G_{\sigma} & M_2+W_{3u} & W_{1s}+W_{3s} & G_{\sigma} & G_{\sigma} & W_{2s} & W_{2s} & G_{\sigma}  \\
    G_{\sigma} & W_{1s}+W_{3s} & M_3+W_{3u} & -G_{\sigma} & G_{\sigma} & W_{2s} & W_{2s} & G_{\sigma} \\
    W_{1s}+W_{3s} & G_{\sigma}& -G_{\sigma} & M_4+W_{3u} & W_{2s} & G_{\sigma} & G_{\sigma} & W_{2s} \\
    W_{2s} & G_{\sigma} & G_{\sigma} & W_{2s} & M_5+W_{3u} & -G_{\sigma}& G_{\sigma} & W_{1s}+W_{3s} \\
    G_{\sigma} & W_{2s} & W_{2s} & G_{\sigma} & -G_{\sigma} & M_6+W_{3u} & W_{1s}+W_{3s} & G_{\sigma} \\
    G_{\sigma} & W_{2s} & W_{2s} & G_{\sigma} & G_{\sigma} & W_{1s}+W_{3s} & M_7+W_{3u} & -G_{\sigma} \\
    W_{2s} & G_{\sigma} & G_{\sigma} & W_{2s} & W_{1s}+W_{3s} & G_{\sigma} & -G_{\sigma} & M_8+W_{3u}
    \end{array}
    \right),
\end{equation}
\begin{equation}
    M_j =
    \left(
    \begin{array}{cc}
              -\dfrac{t_0}{2} [ \cos(k_x+K_{j,x}) + \cos(k_y+K_{j,y}) ]
    & \dfrac{\sqrt{3} t_0}{2} [ \cos(k_x+K_{j,x}) - \cos(k_y+K_{j,y}) ] \\
      \dfrac{\sqrt{3} t_0}{2} [ \cos(k_x+K_{j,x}) - \cos(k_y+K_{j,y}) ]
    &       -\dfrac{3 t_0}{2} [ \cos(k_x+K_{j,x}) + \cos(k_y+K_{j,y}) ]
    \end{array}
    \right),
\end{equation}
\end{widetext}
\begin{eqnarray}
    G_{\uparrow} &=& \left( \begin{array}{cc}
        -J_H S_c/2 & 0 \\
        0 & -J_H S_c/2
    \end{array} \right), \\
    G_{\downarrow} &=& \left( \begin{array}{cc}
        J_H S_c/2 & 0 \\
        0 & J_H S_c/2
    \end{array} \right),
\end{eqnarray}
\begin{eqnarray}
    W_{\rm 1s} &=& \left( \begin{array}{cc}
        -\beta \lambda Q_{\rm 1s} & 0 \\
        0 & -\beta \lambda Q_{\rm 1s}
    \end{array} \right), \\
    W_{\rm 2s} &=& \left( \begin{array}{cc}
        0 & \lambda Q_{\rm 2s} \\
        \lambda Q_{\rm 2s} & 0
    \end{array} \right), \\
    W_{\rm 3u} &=& \left( \begin{array}{cc}
        -\lambda Q_{\rm 3u} & 0 \\
        0 & \lambda Q_{\rm 3u}
    \end{array} \right), \\
    W_{\rm 3s} &=& \left( \begin{array}{cc}
        -\lambda Q_{\rm 3s} & 0 \\
        0 & \lambda Q_{\rm 3s}
    \end{array} \right),
\end{eqnarray}
$Q_{\rm 1s} = 0.053$~{\AA}, $Q_{\rm 2s} = 0.054$~{\AA}, $Q_{\rm 3u} = 0.107$~{\AA}, and
$Q_{\rm 3s} = -0.012$~{\AA} (Ref.~\onlinecite{Herrero11}). The element of the $16 \times 16$ matrix
$H^{\rm dd,HF}_{{\bf k}\sigma}$ is independent of ${\bf k}$,
\begin{eqnarray}
    &&\left(H^{\rm dd,HF}_{{\bf k}\sigma}\right)_{2(j-1)+\xi,2(j'-1)+\xi'}= \nonumber \\
    &&\sum_{{\bf i}_u,\eta}
    \frac{U_{{\bf i}_u\eta\sigma}}{8} e^{-i({\bf K}_{j}-{\bf K}_{j'}) \cdot {\bf i}_u}
    \left( R_{{\bf i}_u\eta} \right)_{\xi \xi'}
\end{eqnarray}
where $\eta=+,-$, $j,j'=1,2,\ldots,8$, and $\xi,\xi'=1,2 $. Further, ${\bf i}_u$ represents
the position of the Mn ions within the unit cell, that is, $(0,0)$, $(a,0)$, $(2a,0)$, $(3a,0)$, $(a,-a)$,
$(2a,-a)$, $(a,a)$, and $(2a,a)$ in Fig.~\ref{fig:ce-order}, and
\begin{eqnarray}
    R_{{\bf i}_u-} &=& \left( \begin{array}{cc}
        \cos^2 \theta_{{\bf i}_u} & \cos \theta_{{\bf i}_u} \sin \theta_{{\bf i}_u} \\
        \cos \theta_{{\bf i}_u} \sin \theta_{{\bf i}_u} & \sin^2 \theta_{{\bf i}_u}
    \end{array} \right), \\
    R_{{\bf i}_u+} &=& \left( \begin{array}{cc}
        \sin^2 \theta_{{\bf i}_u} & -\cos \theta_{{\bf i}_u}\sin \theta_{{\bf i}_u} \\
        -\cos \theta_{{\bf i}_u} \sin \theta_{{\bf i}_u} & \cos^2 \theta_{{\bf i}_u}
    \end{array} \right).
\end{eqnarray}
$\theta_{\bf i}$ is defined from the local lower ($-$) and upper ($+$) Jahn-Teller eigenstate,
\begin{eqnarray}
    d^{\dag}_{{\bf i}-\sigma} &=&
        d^{\dag}_{{\bf i}1\sigma} \cos \theta_{\bf i} +
        d^{\dag}_{{\bf i}2\sigma} \sin \theta_{\bf i}, \\
    d^{\dag}_{{\bf i}+\sigma} &=&
       -d^{\dag}_{{\bf i}1\sigma} \sin \theta_{\bf i} +
        d^{\dag}_{{\bf i}2\sigma} \cos \theta_{\bf i}.
\end{eqnarray}
At Mn$^{3+}$ sites in the $x/y$ directional legs of the zigzag chain in Fig.~\ref{fig:ce-order},
\begin{equation}
    \tan \theta_{\bf i}=\pm \frac{Q_{\rm 3u}+Q_{\rm 3s}-
        \sqrt{(Q_{\rm 3u}+Q_{\rm 3s})^2+4Q_{\rm 2s}^2}}{2Q_{\rm 2s}}
\end{equation}
At Mn$^{4+}$ sites, $\theta_{\bf i}=0$. The matrix for the number operator in reciprocal space is
necessary to evaluate $U_{{\bf i}_u\eta\sigma}$ and its element is given below.
\begin{equation}
    \left(n^{{\bf i}_u\eta\sigma}_{\bf k} \right)_{2(j-1)+\xi,2(j'-1)+\xi'}=
    e^{-i({\bf K}_{j}-{\bf K}_{j'}) \cdot {\bf i}_u}
    \left( R_{{\bf i}_u\eta} \right)_{\xi \xi'}
\end{equation}
We find the eigenstates and eigenenergies of the $16\times16$ matrix
$H^{\rm d}_{{\bf k}\sigma} = H^{\rm d,nonint}_{{\bf k}\sigma} + H^{\rm dd,HF}_{{\bf k}\sigma}$ at chosen set
of ${\bf k}$ points through the Hartree-Fock iterative calculations, which are used to find the electronic DOS
in the absence of the core hole shown in Figs.~\ref{fig:dos-nenh-t0.9}(a) and \ref{fig:dos-nenh-t1.5}(a).

The Hamiltonian in the presence of the core hole at a site ${\bf i}_c$ for $N \times N$ cluster model
of La$_{0.5}$Sr$_{1.5}$MnO$_4$, with $N$ multiple of 4, is presented below.
\begin{eqnarray}
    \hat{H}_{{\rm total},{\bf i}_c} = \sum_{\sigma, \ {\bf k},{\bf k}'\in\Omega_{\rm 1BZ}}
        && d^{\dagger}_{{\bf k} \sigma} \big(
            H^{\rm d,nonint}_{{\bf k} \sigma} \delta_{{\bf k} {\bf k}'} + H^{\rm dd,HF}_{{\bf k} {\bf k}' \sigma} \nonumber \\
        &&  + H^{\rm sd,{\bf i}_c}_{{\bf k} {\bf k}' \sigma}
        \big) \ d_{{\bf k}' \sigma} ,
\end{eqnarray}
where
\begin{eqnarray}
    & \left(H^{\rm dd,HF}_{{\bf k}{\bf k}'\sigma}\right)_{2(j-1)+\xi,2(j'-1)+\xi'} = \nonumber \\
    & \sum_{{\bf i},\eta}
        \frac{U_{{\bf i}\eta\sigma}}{N^2} \
        e^{-i({\bf k}-{\bf k}') \cdot {\bf i}} \
        e^{-i({\bf K}_{j}-{\bf K}_{j'}) \cdot {\bf i}}
        \left( R_{{\bf i}\eta} \right)_{\xi \xi'} ,
\end{eqnarray}
\begin{eqnarray}
    & \left(H^{{\rm sd},{\bf i}_c}_{{\bf k}{\bf k}'\sigma}\right)_{2(j-1)+\xi,2(j'-1)+\xi'} = \nonumber \\
        & U_{\rm core} \ e^{-i({\bf k}-{\bf k}') \cdot {\bf i}_c} \
        e^{-i({\bf K}_{j}-{\bf K}_{j'}) \cdot {\bf i}_c} \ \delta_{\xi \xi'},
\end{eqnarray}
with $\eta =+,-$, $\quad j,j' = 1,2,\ldots,8$, and $\xi,\xi'= 1,2$. For the evaluation of
$U_{{\bf i}\eta\sigma}$, the number operator in receiprocal space is necessary, shown below.
\begin{equation}
    \hat{n}^{{\bf i}\eta\sigma} = \sum_{{\bf k},{\bf k}'\in\Omega_{\rm 1BZ}}
        d^{\dagger}_{{\bf k} \sigma} \ n^{{\bf i}\eta\sigma}_{{\bf k}{\bf k}'} \ d_{{\bf k}' \sigma} ,
\end{equation}
where
\begin{eqnarray}
    &\left(n^{{\bf i}\eta\sigma}_{{\bf k}{\bf k}'}\right)_{2(j-1)+\xi,2(j'-1)+\xi'} = \nonumber\\
    &   e^{-i({\bf k}-{\bf k}') \cdot {\bf i}} \ e^{-i({\bf K}_{j}-{\bf K}_{j'}) \cdot {\bf i}}
        \left( R_{{\bf i}\eta}\right)_{\xi \xi'}.
\end{eqnarray}

Eigenvectors and eigenvalues are found for the $2N^2 \times 2N^2$ matrix of $\hat{H}_{{\rm total},{\bf i}_c}$
for each spin direction $\sigma$ in the presence of the core hole, through the Hartree-Fock iterative
calculations. When necessary, the Pulay mixing method is used to reach a convergence.\cite{Pulay80,Pulay90}
The eigenstates and eigenenergies in the absence of the core hole for the same cluster are found by setting
$U_{\rm core}=0$ and repeating the Hartree-Fock iterative calculations. The two sets of eigenstates and
eigenvalues give $\varepsilon_{l{\bf k}\sigma}$, $\varepsilon_{m\sigma}$, and $\beta_{l {\bf k} m \sigma}$,
which are used for the K-edge RIXS spectrum calculations of La$_{0.5}$Sr$_{1.5}$MnO$_4$.

\section{Electron numbers at Mn$^{3+}$ and Mn$^{4+}$ sites}\label{sec:app-eh-number}

In this Appendix, we discuss the electron numbers on Mn ions. In our effective Hamiltonian, the state
created by $d_{{\bf i}\xi\sigma}^{\dagger}$ is a hybridized state of atomic Mn $3d$  orbital and
surrounding atomic O $2p$ orbitals. Therefore, $e_g$ electron numbers of 0.87 and 0.13 found for
``Mn$^{3+}$" and ``Mn$^{4+}$" sites in our effective Hamiltonian do not represent the actual numbers
on Mn ions, which can be measured by resonant x-ray scattering. For a proper comparison we carry
out an analysis in terms of {\it atomic} Mn $3d$ and O $2p$ orbitals, similar to
Ref.~\onlinecite{vanVeenendaal93}. In the basis of $|d^4\rangle$ and $|d^5\underline{L}\rangle$, where
$d^n$ and $\underline{L}$ represent the presence of $n$ electrons in atomic Mn $3d$ level and a hole
in the ligand O $2p$ level of $e_g$ symmetry, the Hamiltonian for the states with one $e_g$ electron is
\begin{equation}
    H_{{\rm one} \ e_g} = \left(
        \begin{array}{cc}
        0 & 2t_{\rm dp} \\
        2t_{\rm dp} & \Delta \\
        \end{array}
    \right),
\end{equation}
where $t_{\rm dp}$ represents the O $2p$ -- Mn $3d$ electron hopping amplitude and $\Delta$ is the energy
difference between Mn $3d$ and O $2p$ levels. Similarly, the Hamiltonian for the states with zero $e_g$
electron is
\begin{equation}
    H_{{\rm zero} \ e_g} = \left(
        \begin{array}{cc}
        0 & 2t_{\rm dp} \\
        2t_{\rm dp} & \Delta-U_{\rm atomic} \\
        \end{array}
    \right),
\end{equation}
in the basis of $|d^3\rangle$ and $|d^4\underline{L}\rangle$, where $3d$--$3d$ Coulomb interaction $U_{\rm atomic}$
is included to account for one less $3d$ electrons in $|d^4\underline{L}\rangle$ than in $|d^5\underline{L}\rangle$.

The lower energy eigenstates of $H_{{\rm one} \ e_g}$ and $H_{{\rm zero} \ e_g}$,
\begin{eqnarray}
    |{\rm one} \ e_g\ {\rm electron}\rangle &=& \mu_1|d^4\rangle + \nu_1|d^5\underline{L}\rangle , \\
    |{\rm zero}\ e_g\ {\rm electron}\rangle &=& \mu_0|d^3\rangle + \nu_0|d^4\underline{L}\rangle ,
    \\ \nonumber    
\end{eqnarray}
correspond to the states with one and zero electron in hybridized $e_g$ levels considered in our effective
Hamiltonian $\hat{H}_{\rm d}$. Therefore, $|{\rm one}\ e_g\ {\rm electron}\rangle$ state has $4|\mu_1|^2+5|\nu_1|^2$
electrons in atomic Mn $3d$ levels, whereas $|{\rm zero}\ e_g\ {\rm electron}\rangle$ state has
$3|\mu_0|^2+4|\nu_0|^2$ electrons in atomic Mn $3d$ levels. Therefore, $n_{e_g}$ electrons in
the hybridized $e_g$ levels obtained from $\hat{H}_{\rm d}$ corresponds to $n_{\rm atomic\ Mn}$ electrons
on the atomic Mn $3d$ levels, defined as
\begin{eqnarray}
    && n_{\rm atomic\ Mn} = \nonumber \\
    && n_{e_g} (4|\mu_1|^2+5|\nu_1|^2) + (1-n_{e_g}) (3|\mu_0|^2+4|\nu_0|^2).
\end{eqnarray}

For typical values of $t_{\rm dp}$ = 1~eV, $\Delta$ = 4~eV, $U_{\rm atomic}$ = 7~eV, and $n_{e_g}$ = 0.87 and 0.13
obtained for $e_g$ levels around ``Mn$^{3+}$" and ``Mn$^{4+}$" sites for $t_0$ = 0.9~eV case,
we find $n_{{\rm atomic \ Mn}}$ = 4.10 and 3.84 for nominal Mn$^{3+}$ and Mn$^{4+}$ ions,~\cite{Herrero10} and
the difference is only about 0.26, which is much smaller than the difference of 0.74 between $n_{e_g}$'s
and is consistent with 0.15--0.3 suggested by resonant x-ray scattering at the Mn K-edge and bond
valance sum method.~\cite{Herrero11,Brese91}



\begin{thebibliography}{99}
\bibitem{note-JT}
        We note that unlike the three-dimensional manganites, the degeneracy of the $e_g$ levels are
        broken even without the $Q_2$ or $Q_3$ distortions because of the layered crystal structure.
        So, we should consider $\hat{H}_{\rm JT}$ as the effective Hamiltonian that includes not only the effect
        of the oxygen octahedron but also the crystal field of farther ions.
\bibitem{note-PiPi}
        We also note that the Fourier transform of electron number change shown in Fig.~\ref{fig:i-map-t0.9}(a)
        and \ref{fig:i-map-t0.9}(b) has a maximum at $(\pm\frac{\pi}{a},\pm\frac{\pi}{a})$, consistent with
        the point in reciprocal space with maximum integrated RIXS intensity. However, such correspondence
        does not apply to $t_0$ = 1.5~eV case, indicating further study is required to establish the exact
        relation between RIXS intensities in reciprocal space and screening in real space.
\bibitem{note-beta}
        In other words, from the $N_e/2 \times N_e/2$ part of the matrix of $\beta_{l{\bf k}m\sigma}$ with
        $m=1,\ldots,N_e/2$ and $\varepsilon_{l{\bf k}\sigma} < \varepsilon_{\rm F}$, the row corresponding to
        ${\bf k}={\bf k}_h$ and $l=l_h$ is replaced by the row corresponding to ${\bf k}={\bf k}_e$ and $l=l_e$
        in the matrix of $\beta_{l{\bf k}m\sigma}$.
\end{thebibliography}
\end{document}